\documentclass[10pt,prl,aps,twocolumn,floatfix,preprintnumbers,superscriptaddress, nofootinbib]{revtex4-2}
 \usepackage{tikz,cancel}
   \usepackage[normalem]{ulem}
   \usetikzlibrary{shapes.geometric,arrows.meta,positioning}
\usepackage{amsfonts,amssymb,physics}
\usepackage{stmaryrd}
\usepackage{mathtools}
\usepackage{lmodern}
\usepackage{CJKutf8,enumitem}
\usepackage[T1]{fontenc}
\usepackage[utf8]{inputenc}
\usepackage{graphicx}
\usepackage{comment}
\usepackage{bbm,soul}
\usepackage{mathrsfs}

\DeclareMathAlphabet{\mathbfi}{OML}{cmm}{b}{it}
\usepackage{tikz}
\definecolor{mLightBrown}{HTML}{EB811B}
\usetikzlibrary{shapes.geometric, arrows}
\tikzstyle{startstop} = [rectangle, rounded corners, 
minimum width=1cm, 
minimum height=1cm,
text centered, 
draw=black,
fill=pink!50!yellow!90!orange]

\tikzstyle{decision} = [diamond, 
minimum width=2.85cm, 
minimum height=2.92cm, 
text width=2cm, 
text centered, 
draw=teal!70!green, 
fill=teal!70!green!60]

\tikzstyle{process} = [rectangle, rounded corners, 
minimum width=1.5cm, 
minimum height=1.05cm, 
text centered, 
text width=1.8cm, 
draw=mLightBrown, 
fill=mLightBrown!40!white]

\tikzstyle{arrow} = [thick,black!70,-latex,>=stealth]

\usepackage[colorlinks=true,
  linkcolor=blue,
  citecolor=blue,
  urlcolor=blue]{hyperref}
\usepackage{xcolor}

\definecolor{darkpastelgreen}{rgb}{0.01, 0.75, 0.24}

\newcommand{\spacetime}{\ensuremath{\mathcal{M}}}
\newcommand{\ospace}{\ensuremath{\mathcal{O}}}
\newcommand{\fspace}{\ensuremath{\mathcal{F}}}

\newcommand{\fggroup}{\ensuremath{\mathcal{G}}}
\newcommand{\physf}{\ensuremath{\hat\fspace}}
\newcommand{\diff}{\mathrm{d}}
\newcommand{\vfields}{\ensuremath{\mathfrak{X}}(\fspace)}
\newcommand{\fundvfields}{X}
\newcommand{\fmetric}{\ensuremath{\mathfrak{G}}}
\newcommand{\invfmetric}{\ensuremath{\bar{\mathfrak{G}}}}
\newcommand{\gf}[1]{\ensuremath{\tilde{#1}}}

\begin{document}

\title{
{Relational path integral, effective actions and quantum frame covariance in gravity }}

\author{Sergio E. Aguilar-Gutierrez}
\email{sergio.ernesto.aguilar@gmail.com}
\affiliation{Qubits and Spacetime Unit, Okinawa Institute of Science and Technology Graduate University (\begin{CJK}{UTF8}{min}沖縄科学技術大学院大学\end{CJK}), Onna, Okinawa 904 0495, Japan}

\author{Renata Ferrero} 
\email{renata.ferrero@fau.de}
\affiliation{Institute for Quantum Gravity, Friedrich-Alexander-Universit{\"a}t Erlangen-N{\"u}rnberg, Staudtstr. 7, 91058 Erlangen, Germany}

\author{Philipp A. Höhn} 
\email{philipp.hoehn@oist.jp}
\affiliation{Qubits and Spacetime Unit, Okinawa Institute of Science and Technology Graduate University (\begin{CJK}{UTF8}{min}沖縄科学技術大学院大学\end{CJK}), Onna, Okinawa 904 0495, Japan}

\author{Luca Marchetti} 
\email{luca.marchetti@oist.jp}
\affiliation{Kavli Institute for the Physics and Mathematics of the Universe (WPI),\\ UTIAS, The University of Tokyo, Chiba 277-8583, Japan}
\affiliation{Qubits and Spacetime Unit, Okinawa Institute of Science and Technology Graduate University (\begin{CJK}{UTF8}{min}沖縄科学技術大学院大学\end{CJK}), Onna, Okinawa 904 0495, Japan}

\date{\today}

\begin{abstract}
We propose a relational bundle-geometric formulation of the gravitational path integral by invoking the new tool of quantum reference frames (QRFs), which in gravity are gauge-covariant coordinate systems constructed from the available field content. Formulated in terms of relational (frame-dressed) observables, this yields a manifestly gauge-invariant path integral without ghosts and anomalies, and in which observables and their correlators are local to a frame. While eliminating the need for gauge fixing, it is equivalent to Faddeev-Popov versions in which the QRF is gauge-fixed, recovering certain previous proposals. A key feature is its covariance under QRF changes: it is a perspective-neutral path integral which encodes all internal QRF perspectives and the transformations between them. This leads to several qualitative predictions: local correlators and time evolution of relational observables in one QRF perspective become `fuzzy’ in another, and a new spectrum of relational vacua arises. Comprised of frame-dependent no-boundary and asymptotic ground states,  a vacuum from one perspective appears generally `excited’ in another. Finally, we construct gauge-invariant, yet frame-dependent effective actions by coupling sources exclusively to relational observables, setting the stage for a relational definition of renormalization.
\end{abstract}

\maketitle

The gravitational path integral \cite{Hawking:1978jz,Hawking:1980gf} is a central object in covariant approaches to quantum gravity \cite{Loll:2019rdj,Ambjorn:2012jv,Ambjorn:2026prt,green2012superstring,polchinski1998string,Penedones:2016voo,hartman2015lectures,Perez:2012wv,rovelli2015covariant,Krajewski:2011zzu,gielen2020progress,Oriti:2017ave,Percacci:2017fkn,Reuter:2019byg,Eichhorn:2026uqj,Ferrero:2025efd}. It defines transition amplitudes, correlation functions,  physical inner products \cite{Halliwell:1990qr,Rovelli:1998dx}, and effective actions. The latter encode how quantum fluctuations modify mean field dynamics and provide the starting point for renormalization group analyses \cite{Percacci:2017fkn,Reuter:2019byg,Eichhorn:2026uqj,Ferrero:2025efd}. In Euclidean signature, it further defines vacuum structures \cite{Hartle:1983ai,Lehners:2023yrj,Maldacena:2024uhs},
and offers computational tools for entropies \cite{Gibbons:1976ue,Gibbons:1977mu,Susskind:1994sm,Lewkowycz:2013nqa,Rangamani:2016dms,Colafranceschi:2023moh}, the Page curve \cite{Penington:2019kki,Almheiri:2019hni,Almheiri:2019qdq,Penington:2019npb,Marolf:2020rpm,Chowdhury:2021nxw}, closed universe puzzles \cite{Marolf:2020xie,Usatyuk:2024mzs,Abdalla:2025gzn,Harlow:2025pvj,Chen:2025fwp,Abdalla:2026mxn,Zhao:2026mpl}, and more generally the `factorization problem' in holography \cite{Maldacena:2004rf,Liu:2025ikq}.

Despite its central role, the gravitational path integral remains plagued by difficulties, many of which can be traced back to diffeomorphism invariance. For example, at the technical level, gauge-invariant formulations 
need non-anomalous measures, 
the Gribov problem obstructs a global gauge fixing, while a proper definition requires  regulators which typically break diffeomorphism invariance. At the conceptual level, spacetime transition amplitudes yield the physical inner product \cite{Halliwell:1990qr,Rovelli:1998dx}, not time evolution. Correlation functions cannot be ordinary $n$-point functions of spacetime-local fields since these are not gauge-invariant, and effective actions encoding these are typically either not fully diffeomorphism-invariant \cite{DeWitt:1967ub,DeWitt:1967yk,Abbott:1981ke,
Abbott:1980hw,Niedermaier:2006wt,casadio2023background,Falls:2025tid}, or involve non-local  fields without clear physical interpretation \cite{DeWitt:1980jv,Vilkovisky:1984st,Burgess:1987zi,Huggins:1987zw,Lavrov:1988is,gospel,DeWitt:2003pm}. And regarding renormalization, what meaning does scale have without a background?

Here, we revisit these challenges from a novel geometric and relational perspective, pairing DeWitt's old bundle approach \cite{DeWitt:2003pm,Parker:2009uva,DeWitt:1998eq,Kunstatter:1991ds} with the modern tool of quantum reference frames (QRFs) \cite{Goeller:2022rsx,Carrozza:2022xut,Freidel:2025ous,Freidel:2026stu,Thiemann:2026vaj,Marchetti:2024nnk,Chen:2026kui,DeVuyst:2024pop,DeVuyst:2024uvd,delaHamette:2021oex,Hoehn:2023ehz,Hoehn:2019fsy,AliAhmad:2021adn,delaHamette:2026rtv,Araujo-Regado:2025ejs,Giacomini:2021gei,Castro-Ruiz:2019nnl,Apadula:2022pxk,Giacomini:2017zju,Garmier:2025soc,Fewster:2024pur,Carette:2023wpz,delaHamette:2020dyi} to deal with diffeomorphisms. The result is a manifestly gauge-invariant \emph{relational path integral} without ghosts and anomalies, which, when gauge-fixed, recovers the recent proposal \cite{Falls:2025tid}. Encoding all QRF perspectives, it also constitutes a path integral and gravity incarnation of the perspective-neutral approach to QRF covariance \cite{delaHamette:2021oex,Hoehn:2023ehz,Hoehn:2019fsy,AliAhmad:2021adn,DeVuyst:2024pop,DeVuyst:2024uvd,delaHamette:2026rtv,Araujo-Regado:2025ejs,Giacomini:2021gei,Chen:2026kui,Castro-Ruiz:2019nnl,Apadula:2022pxk}.

Key to addressing the challenges is that each frame `sees' spacetime (locally) in a `specially covariantized' form, called \emph{relational spacetime}; global coverage is achieved via a `meta-atlas' of frames. In relational spacetime, correlators of relational observables become ordinary $n$-point functions, encoded in gauge-invariant generating functionals and effective actions, transition amplitudes encode bulk time evolution, and one may regularize without breaking gauge invariance. The resulting description is  \emph{frame-dependent}, yet \emph{frame-covariant}.
Additionally, our effective actions give rise to a new relational notion of renormalization and scale \cite{AFHM1,AFHM2}.

Our construction proceeds at a non-perturbative, so somewhat symbolical level of quantum gravity (fixed topology for now). It is approach-independent  and may be regarded as both a `road map' for explicit realizations, and a new paradigm for the gravitational path integral. Moreover,  the symbolical nature still permits us to derive \emph{qualitative} physical predictions. 
We show how sharp correlators and time evolutions in one frame perspective become fuzzy in another, in generalization of `event relativity' \cite{Nitti:2024iyj,Kabel:2024lzr}. We will also encounter \emph{relational Hartle-Hawking states} and asymptotic vacua of bulk Hamiltonians in relational spacetime; what is a ground state for one frame, appears generically excited  to another.

\paragraph{Field space.}
We consider generally covariant theories whose field space $\fspace$\,--\,the set of kinematical field configurations $\phi$ on some spacetime manifold $\spacetime$ (of fixed topology)\,--\,constitutes a principal fiber bundle. Its typical fiber $\fggroup\subset \mathrm{Diff}(\spacetime)$ is given by `small diffeomorphisms' and base $\physf\coloneqq\fspace/\mathcal{G}$ carries the gauge-invariant information (see App.~\ref{app:geometry} for details).

To define a gauge-invariant path integration, DeWitt formally introduced an invariant metric $\fmetric$ and `fiber-adapted coordinates' on $\fspace$ that split gauge from gauge-invariant data \cite{DeWitt:2003pm,DeWitt:1967ub,DeWitt:1967yk,DeWitt:1980jv,Parker:2009uva,Kunstatter:1991ds,DeWitt:1998eq, gospel}. However, the spacetime non-locality of  gauge-invariant coordinates and ensuing difficulties in imposing boundary and causality conditions in $\spacetime$ led to these ideas later being abandoned \cite{DeWitt:2003pm,Pawlowski:2003sk,Donkin:2012ud}. Here, we aim to revive them by invoking dynamical frames to address these challenges \emph{relationally}.

\paragraph*{{Dynamical reference frames.}} {Dynamical frames constitute the degrees of freedom that are either gauge-fixed or used as dressing when extracting physical information from a gauge system, yielding a split between gauge-redundant (the frame) and gauge-invariant data. We will adopt the  generally covariant framework  for dynamical frames developed in \cite{Goeller:2022rsx,Carrozza:2022xut,Freidel:2025ous} (see \cite{Gomes:2021tjm,Gomes:2024zgy,Francois:2024rdm,Giddings:2025xym,Giddings:2025bkp,Thiemann:2026vaj,Marchetti:2024nnk,Cheung:2026euf,Chen:2026kui,Wei:2025guh} for somewhat related work), which unifies and generalizes the dressed \cite{Donnelly:2015hta,Donnelly:2016rvo,Giddings:2025xym,Cheung:2026euf,Harlow:2021dfp,Blommaert:2019hjr,Heemskerk:2012np,Kabat:2013wga} and relational  \cite{Bergmann:1960wb,Bergmann:1961wa,Rovelli:1990ph,Rovelli:1990pi,Rovelli:2004tv,Giddings:2005id,Dittrich:2004cb,Dittrich:2005kc,Tambornino:2011vg,Giesel:2007wi,Brunetti:2013maa,Brunetti:2016hgw,Khavkine:2015fwa,Frob:2017lnt,Baldazzi:2021fye,Frob:2022ciq} approaches to gravitational observables. The frame fields are given by \emph{field-dependent} gauge-covariant coordinate systems $R[\phi]:\mathcal{M}\supset \mathcal{U}_{R}{[\phi]}\to \mathcal{U}_{\ospace}\subset\ospace_R$, mapping field-dependent spacetime regions to regions in \emph{relational spacetime} $\ospace_R$ and transforming as a collection of $\dim\spacetime$ scalars $R[f_\star {\phi}] =f_\star R[\phi]= R{[\phi]} \circ f^{-1}$, $f\in\fggroup$.  
They are to $\mathcal{G}$ what tetrads are to the Lorentz group (see App.~\ref{app_frames}). 
As with ordinary coordinates, there exists a plethora of ways to construct dynamical frames from the given field content, e.g.\ via geodesic or matter dressings \cite{Goeller:2022rsx,Carrozza:2022xut}, Bergmann scalars \cite{Bergmann:1960wb,Bergmann:1961wa}, Fermi coordinates, etc.  }

Given any spacetime-local {gauge-covariant quantity $T[f_\star{\phi}]=f_\star T[\phi]$, $f\in\mathcal{G}$  (e.g., some tensor field), a corresponding gauge-invariant frame-dressed observable is obtained by pushing $T$ forward to $\ospace_R$ with frame $R$ \cite{Goeller:2022rsx},
\begin{equation}\label{eq:relational1}
    O_{T\mid R}[{\phi}] \equiv (R[{\phi}])_\star T[{\phi}]\,,
\end{equation}
 where $O_{T|R}$ is \emph{local} (see App.~\ref{app_frames} for details and \cite{Carrozza:2022xut,Baldazzi:2021fye} for dressing non-covariant quantities). This is a \emph{relational observable}, measuring the components of $T$, where $R$ reads $R[\phi](x)=o\in\ospace_R$. 

If $g$ is the metric on $\spacetime$, $O_{g|R}$ is a metric on $\ospace_R$ of the same signature.  $(\ospace_R,O_{g|R})$ is the \emph{physical} spacetime `as seen' by $R$. Key to addressing DeWitt's challenges is to note that this is the home of relational observables, where they are local, not $\spacetime$, where they are non-local. Indeed, it is $\ospace_R$, where propagation and microcausality \cite{Goeller:2022rsx,Marolf:2015jha,Hoehn:2025pmx} (at least for $\spacetime$-locally constructed frames), and boundary conditions can be locally defined.

As shown in Apps.~\ref{app:geometry} and~\ref{app_metaatlas}, each frame $R[\phi]$ locally
provides a horizontal connection, as well as a \emph{relational fiber-adapted coordinate system} $(R^\alpha,O^A_R)$ ($\alpha,A$ are DeWitt indices). This construction
is local in \emph{both} $\fspace$ and $\spacetime$; covering both therefore requires a `meta-atlas' of frames. The frame variables $R^\alpha$ coordinatize the  fibers, while the associated relational observables $O^A_R$ coordinatize the $R$-horizontal directions. Thanks to gauge invariance, the $O^A_R$ descend to local coordinates $\hat O^A_R$ on $\physf$.

We distinguish three types of transformations involving frames (cf.~Apps.~\ref{app_frames} and~\ref{app:geometry}).  \emph{Frame changes} change the description, but not the field configuration, come in two  guises, and yield a \emph{physical} frame covariance \cite{Goeller:2022rsx}. First, given two frames $R,R'$, we have gauge-invariant \emph{spacetime} coordinate transformations 
\begin{equation}\label{eq:frame-change-map}
    U_{R\rightarrow R'}[\phi]
    =
    R'[\phi]\circ R^{-1}[\phi]=O_{R|R'}[\phi]:\ospace_R\to\ospace_{R'}\,,
\end{equation}
changing relational spacetime and coinciding with the relational observables describing $R'$ relative to $R$. Relational observables transform appropriately 
\begin{equation}
    O_{T|R'}[\phi]
    =
    \big(U_{R\rightarrow R'}[\phi]\big)_\star
    O_{T|R}[\phi]\,.
    \label{eq:frame-change}
\end{equation}
This is an active \emph{change of observables}.
Second, mapping histories between relational spacetimes, Eq.~\eqref{eq:frame-change} induces \emph{base space} coordinate transformations $\hat O^{A'}_{R'}[\hat O^A_R]$. This leads to a passive \emph{change of description} of the same observables on $\physf$.

Third, \emph{frame reorientations}  change the field configuration, but not the description \cite{Carrozza:2022xut,Goeller:2022rsx,Freidel:2025ous}. Given frame $R$, these are active field-independent diffeomorphisms in $\ospace_R$ (acting only on dynamical fields in $\ospace_R$). These are \emph{physical}; boundary symmetries  \cite{Carrozza:2022xut,Carrozza:2021gju,Araujo-Regado:2024dpr,Araujo-Regado:2025ejs}, and relational evolution below furnish examples.

\paragraph*{Relational path integral.}
The metric $\fmetric$ induces a metric $\gamma$ on the gauge-orbits  (assumed invertible) and a volume form $\hat{\epsilon}$ on $\physf$, producing a
measure on $\physf$ (App.~\ref{app:geometry}):
\begin{equation}\label{eq:basespacemeasure}
\mathcal{D}\mu[\hat O_R]=\hat{\epsilon}\times[\det\gamma]^{1/2}[\hat{O}_R]\,.
\end{equation}
Crucially, it is covariant under arbitrary frame changes and reorientations\,--\,a key result for frame covariance below. Moreover, it is gauge-invariant and so non-anomalous (an anomaly requires a non-invariant $\fmetric$, cf.~App.~\ref{app:geometry}).

   We may now formally define a gauge-invariant path integral.  Integrating an invariant functional $F[O_R]$ over $\fspace$, results in a (divergent) $\mathcal{G}$-volume contribution. 
   Dropping this redundant factor yields a  relational path integral that is formally finite and equivalent to the base integral (App.~\ref{app:geometry}):
\begin{equation}
    Z
    =
\int_{\hat\fspace}\hat{\epsilon}\,\sqrt{\det\gamma[\hat\phi]}\,e^{iS[\hat\phi]}=
    \int_{\hat\fspace} \mathcal{D}
    \mu_R[\hat{O}_R]\,e^{iS_R[\hat{O}_R]}\,,
    \label{eq:relational-path-integral}
\end{equation}
where $\hat\phi\in\physf$, and the classical action translates as $S[\hat\phi]=\int_\spacetime L= \int_{\ospace_R}R_\star L=S_R[\hat{O}_R]$ for an $\spacetime$-global frame; in general, both $Z,S_R$ must be `patched up' with a meta-atlas of frames, App.~\ref{app_metaatlas}. 
As in \cite{Freidel:2025ous,Freidel:2026stu}, the relational description leads to a non-anomalous quantum theory. 

Eq.~\eqref{eq:relational-path-integral} has several advantages. First, the left expression manifests that $Z$ is \emph{geometrically defined} and so independent of  coordinates on $\physf$. Indeed, $\det\gamma,S$ are scalars (App.~\ref{app:geometry}), so  $Z$ is the integral of a covariant top form. In practice, one evaluates it in specific coordinates, as on the right, where the action is written as a relational spacetime integral $S_R$. As a shorthand, we henceforth write the path integral using a single frame label $R$, keeping in mind, however, that in general it stands for an entire meta-atlas of frames covering both $\spacetime$ and $\physf$ (cf.~App.~\ref{app_metaatlas}).

Second, $Z$ is manifestly gauge invariant, so no ghost sector or gauge fixing is needed, cf.\ \cite{DeWitt:1998eq}. 
Gauge directions have not been fixed but divided out.  Before descending to the  base integral, one may nevertheless fix the frame configuration via $\chi_{R,\gf{R}}[\phi]=R[\phi]-\gf{R}\overset{!}{=}0$, obtaining the equivalent Faddeev-Popov gauge-fixed $\fspace$-integral 
\begin{align}\label{eq:Z_relational}
Z&=\int_{\fspace}\mathcal{D}{R}\,\delta[\chi_{R,\gf{R}}[\phi]]\,\mathcal{D}\mu_R[{O}_R]\,e^{iS_R[{O}_R]}\nonumber\\
    &=\int_\fspace\mathcal{D}{\varphi}\sqrt{\det\fmetric}\,\delta[\chi_{R,\gf{R}}[\varphi]]\det \Delta_{R}\, e^{iS[\varphi]}\,,
\end{align}
where $\varphi$ denotes the original field coordinates on $\fspace$ to which the measure in Eq.~\eqref{eq:basespacemeasure} ascends (App.~\ref{app:geometry}). Thus, the relational path integral is a gauge-invariant parent formulation whose gauge-fixed representations can be expressed either in terms of the $\varphi$-coordinates (second line), with the Faddeev--Popov determinant $\det \Delta_R$
producing ghosts \cite{Faddeev:1967fc,Faddeev:1969su}, or in ghost-free relational fiber-adapted  coordinates (first line). Upon a variable change, Eq.~\eqref{eq:Z_relational} also recovers 
the  proposal in \cite{Falls:2025tid} (see App.~\ref{sapp:dressed field functional integral}).

Third, as an integral over relational observables $\hat{O}_R^A$, it can be regularized directly on $\ospace_R$, rather than  $\spacetime$. Accordingly, one may introduce regulators that are independent of any spacetime background structure, and spacetime diffeomorphism invariance can be preserved at the level of the regularized path integral. The regulators will be frame-dependent, however, as in  \cite{Freidel:2026stu}. 

Fourth, $Z$ opens novel avenues for explicit perturbative constructions. Besides standard perturbations around a spacetime background solution of the fields $\varphi$, one may perturb around a background solution of $\hat O^A_R$ in $\ospace_R$, thereby preserving manifest gauge invariance and possibly sidestepping linearization instabilities \cite{DeVuyst:2024grw,Moncrief:1978te,Higuchi:1991tk,Higuchi:1991tm,Arms:1982ea}.

Finally, since $Z$ is invariant under arbitrary field redefinitions on $\physf$, it is in particular \emph{frame-independent}, i.e.\ invariant under frame changes $\hat{O}^{A'}_{R'}[\hat{O}^A_R]$. (When non-trivial integration limits are imposed, these must be transformed as well, see below.) Nevertheless, $Z$ encodes the information about every internal frame perspective and the transformations among them. As such, $Z$ constitutes a path integral formulation of the \emph{perspective-neutral} framework of QRF covariance \cite{delaHamette:2021oex,Hoehn:2023ehz,Hoehn:2019fsy,AliAhmad:2021adn,DeVuyst:2024pop,DeVuyst:2024uvd,delaHamette:2026rtv,Araujo-Regado:2025ejs,Giacomini:2021gei,Chen:2026kui,Castro-Ruiz:2019nnl,Apadula:2022pxk} for gravitational theories, which we now explore.

\paragraph*{Sharp and fuzzy correlators.} 
Given a collection of relational observables,  their $n$-point
functions read
\begin{align}
   & \left\langle
        \hat{O}_{T_1|R}(o_1)\cdots \hat{O}_{T_n|R}(o_n)
    \right\rangle_R
    =    \label{eq:relational-n-point}\\&
    \frac{1}{Z}
    \int \mathcal{D}
    \mu_R[\hat O_R]\,
    \hat{O}_{T_1|R}(o_1)\cdots \hat{O}_{T_n|R}(o_n)\,
    e^{i S_R[\hat O_R]} ~,\nonumber
    \end{align}
leaving state information, residing in contour and boundary conditions, implicit. This is a standard correlator, defined in $\ospace_R$ at \emph{sharp}, i.e.\ field-independent, events $o_i$. 

To understand how a second frame $R'$  `sees' the same correlator (App.~\ref{app_fuzzy}), we first clarify how to rewrite the $\hat{O}_{T_i|R}$ as  local observables on $\ospace_{R'}$:
\begin{equation}\label{eq:relobsrewrite}
    \hat{O}_{T_i|R}(o_i)=\hat{O}_{\tilde{T}_i|R'}(o_i'[\hat\phi])\,,\quad\hat{O}_{\tilde{T}_i|R'}\coloneqq \hat{O}_{T_i|R}\circ \hat{U}_{R'\to R}\,,
\end{equation}
where $o_i'[\hat\phi]\coloneqq \hat U_{R\to R'}[\hat\phi](o_i)$, with $\hat{U}_{R\to R'}=\hat{O}_{R'\mid R}$,  is the \emph{field-dependent} event in $\ospace_{R'}$ corresponding to $o_i\in\ospace_R$, see Fig.~\ref{fig:sharp_vs_fuzzy}. $\hat{O}_{\tilde{T}_i|R'}$ is a collection of \emph{scalars} on $\ospace_{R'}$, assigning each component of $\hat{O}_{T_i|R}$ to \emph{distinct} events  $o'_i[\hat\phi]$ in $\ospace_{R'}$ for different $\hat\phi$. Second, $\hat{O}_{\tilde{T}_i|R'}(o'_i[\hat\phi])=F[\hat{O}_{R'}^{A'}]$ can be entirely written as functionals of the $R'$ relational coordinates. 

Hence, Eq.~\eqref{eq:relational-n-point} equals 
\begin{align}
   &  \left\langle
        \hat{O}_{\tilde{T}_1|R'}(o'_1[\hat\phi])\cdots \hat{O}_{\tilde{T}_n|R'}(o'_n[\hat\phi])
    \right\rangle_{R'}  = \label{eq:fuzzycorrelator}\\&
    \frac{1}{Z}
    \int \mathcal{D}
    \,\mu_{R'}[\hat O_{R'}]\,
    \hat{O}_{\tilde{T}_1|R'}(o'_1[\hat\phi])\cdots \hat{O}_{\tilde{T}_n|R'}(o'_n[\hat\phi])\,
    e^{iS_{R'}}, \nonumber
    \end{align}
where $\hat{O}_{\tilde{T}_i|R'}(o'_i[\hat\phi])=F[\hat{O}_{R'}^{A'}]$ is understood. The result is a correlator in $\ospace_{R'}$ defined at field-dependent and so, through the path integration, generally \emph{fuzzy} events $o_i'[\hat\phi]$. This is a manifestation and generalization of the observation  \cite{Nitti:2024iyj,Kabel:2024lzr} that the sharpness of an event in quantum gravity is QRF-relative, which will accompany us in the remainder, see Fig.~\ref{fig:sharp_vs_fuzzy}. 

\begin{figure}[h!]
    \centering
    \includegraphics[width=0.68\linewidth]{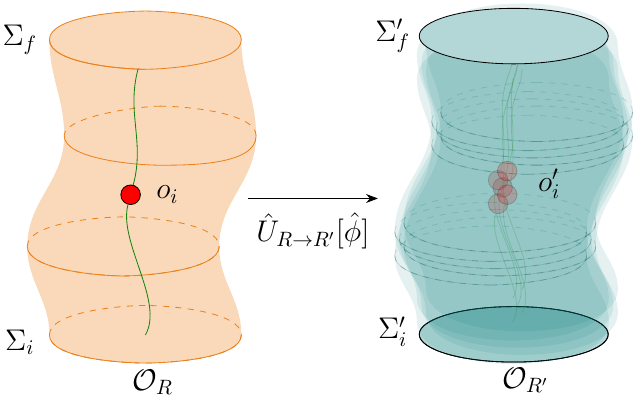}
    \caption{
    A sharply localized event $o_{i}$ (red) in a foliation between two Cauchy slices  $\Sigma_{i,f}$ in $\ospace_R$   (left) corresponds to a `fuzzy' (field-dependent) event $o_{i}'$ in a `fuzzy' foliation between sharp  slices $\Sigma'_{i,f}$ in $\ospace_{R'}$ (right).}
    \label{fig:sharp_vs_fuzzy}
\end{figure}
Let us expound on this using generating functionals. Since $Z$ is formulated directly in terms of gauge-invariant variables, we may define \emph{relational Schwinger functionals} by coupling sources directly to relational observables:
\begin{equation}
    Z_R[J_R]\coloneqq e^{i W_R[J_R]}=\!\int_{\physf} \mathcal{D}
    \mu_R[\hat O_R]\,e^{i
            S_R[\hat O_R]
            +
            i(J_R)_{A}\hat O_R^A}.\label{eq:relational-generating-functional}
\end{equation}
The pairing $(J_R)_{A}\hat O_R^A$ denotes DeWitt contraction in $\ospace_R$. Crucially, the sources $\{J_R\}$ are manifestly gauge-invariant objects, defined as background fields on relational spacetime $\ospace_R$, and therefore intrinsically tied to the chosen frame $R$; i.e.~$J_{R|R'}[\hat\phi]=(\hat{U}_{R\to R'}[\hat\phi])_\star J_R$ are no longer background fields under  frame changes in $\ospace_{R'}$. Hence, the generating functionals are not covariant under frame changes, $Z_R[J_R]\neq Z_{R'}[J_{R'}]$.

Restricting to correlators for the  relational coordinates, $\{\hat{O}_{T_i|R}\}=\{\hat O^{A_i}_R\}$, Eqs.~\eqref{eq:relational-n-point} and~\eqref{eq:fuzzycorrelator} lead to
\begin{align}
& \frac{(-i)^n}{Z_R}\frac{\delta^n Z_R[J_R]}{\delta J_R(o_1)\cdots\delta J_R(o_n)}\Bigg|_{J_R=0}\!\!\!=\!\left\langle\!
        \hat{O}^{A_1}_R(o_1)\cdots \hat{O}^{A_n}_{R}(o_n)\!
    \right\rangle_R \nonumber\\
& =\left\langle
        \hat{O}_{\tilde{A}_1|R'}(o'_1[\hat\phi])\cdots 
    \right\rangle_{R'}\neq \frac{(-i)^n}{Z_{R'}}\frac{\delta^n Z_{R'}[J_{R'}]}{\delta J_{R'}(o'_1)\cdots}\Bigg|_{J_{R'}=0}.\label{eqn:correlators}
\end{align}
Thus, the fuzziness of the correlator in the $R'$-perspective prevents it from being written via derivatives of the $R'$-generating functional, in line with the observation that $Z_R[J_R]$ is not a scalar under frame changes. 

Thus, a frame change affects only the \emph{representation} of the 
correlator, not  its gauge-invariant physical information. However, the `sharpness' of an $n$-point function is QRF-relative; a
correlator sharply localized at $o_1,\cdots,o_n\in\mathcal{O}_R$ generally turns into state-dependent smeared insertions in $\ospace_{R'}$. This discussion also clarifies that descriptions of the same correlator relative to distinct frames can generally not be related by the action of  effective diffeomorphisms\,--\,unless the state is sufficiently semiclassical\,--\,as this would map sharp correlators into sharp correlators.

\paragraph*{Relational effective action.}
Given Eq.\ \eqref{eq:relational-generating-functional}, we now define a \emph{relational} effective action by
\begin{equation}
    \Gamma_R[\bar{O}_R]
    \coloneq
    W_R[J_R]
    -
    (J_{R})_{A}\bar{O}_R^A \,,
    \quad
    \frac{\delta\Gamma_R}{\delta \bar{O}_R^A}
    =
    -(J_{R})_{A}\,,
    \label{eq:relational-effective-action}
\end{equation}
where the mean value of a relational observable is $\bar{O}_R^A=\delta W_R[J_R]/\delta(J_R)_A$. Crucially, this effective action is manifestly gauge invariant and background-independent. Moreover, it computes $1$PI $\ospace_R$-local correlators of relational
observables, and therefore carries a direct physical interpretation, while still encoding the information about any compatible frame perspective, cf.~App.~\ref{app:relea}. In
this sense, it overcomes the typical difficulties associated with defining effective actions in gravitational theories discussed in App.~\ref{app:EA_comparison}. Much
like $Z_R$ in Eq.~\eqref{eq:relational-generating-functional}, the effective action is in general \emph{not} frame-covariant, since generating functionals
associated with different frames generate correlation functions of different relational observables. 
Accordingly, a transformation $\Gamma_R\to \Gamma_{R'}$
 may be interpreted as motion in theory space, and, as explained in App.~\ref{app:relea} via a `frame-Nielsen identity', is parametrized by inessential couplings and generated infinitesimally by redundant operators \cite{Wegner:1974sla,Dietz:2013sba, Baldazzi:2021ydj,Baldazzi:2021orb,Falls:2025sxu,Kuntz:2026vhs,Falls:2026nuh}.

For each frame $R$, the effective action defines a quantum stationary locus through $\delta \Gamma_{R}/\delta\bar{O}_R^A=0$, or equivalently $J_R=0$. We refer to the frame-indexed collection of these loci as the \emph{quantum shell}, which provides a frame-neutral criterion for on-shellness. On-shell frame covariance can then be formulated as the requirement that the background field transformations relating sources in different frames\,--\,which, as noted, cannot be related by $\hat U_{R\to R'}$\,--\,obeys $J_R=0\Longrightarrow J_{R'}=0$ for any $R$ and $R'$, see App.\ \ref{app:relea}.
On shell, $\Gamma_R=-i\ln Z$, so $\Gamma_R$ is frame-covariant and transforms as a scalar.

\paragraph*{Evolution and frame change kernels.}
Consider field-independent initial  and final Cauchy slices $\Sigma_i,\Sigma_f$ in $\ospace_R$, on which we fix the relational field configurations to $\hat{O}^A_{R}[\hat\phi]\big|_{\Sigma_{n}}\equiv{Q}_{R}^{A,n}$, $n=i,f$, respectively. We  assume that this restricts $\Sigma_{i,f}$ to be spacelike. The corresponding path integral kernel reads
\begin{equation}\label{eqn:path-integralkernel0}
K\big[{Q}^{A,f}_{R}, \Sigma_f;{Q}_{R}^{B,i},\Sigma_i\big]  \coloneqq\int_{\hat{\fspace}_{R}^{if}}\mathcal{D} \mu_R[\hat{O}_{R}]\,e^{iS_{R}[\hat{O}_{R}]},
\end{equation}
where $\hat{\fspace}_R^{if}\subset\physf$ is the (restricted) history subspace.

Let us translate this kernel into a second frame $R'$'s perspective. Besides the integrant, we must also transform the parametrization of $\hat{\mathcal{F}}^{if}_R$ in the arguments of $K$.
Crucially, the images of the bounding Cauchy slices $\Sigma'_{i,f}\coloneqq\hat{U}_{R\to R'}(\Sigma_{i,f})$ in $R'$-spacetime $\ospace_{R'}$ are field-independent, because ${\hat U_{R\to R'}=\hat O_{R'|R}}$ is one of the fixed observables on $\Sigma_{i,f}$, and spacelike. Any field-independent slices or events in-between $\Sigma_{i,f}$ become fuzzy in $\ospace_{R'}$, however, cf.~Fig.~\ref{fig:sharp_vs_fuzzy}.
Fixing $\hat O^{A}_{R}$ on $\Sigma_{i,f}$ means fixing the corresponding scalarized observables from Eq.~\eqref{eq:relobsrewrite} to the same configuration, denoted $\tilde Q^{A}_{R'}$, on $\Sigma'_{i,f}$ (fixing also  $\hat O^{A'}_{R'}$).
Whence,
\begin{align}\label{eq:kerneltranslation}
    K\!\big[{Q}_{R}^{A,f}, \Sigma_f;{Q}_{R}^{B,i},\Sigma_i\big]\!=\!K\!\bigg[\tilde{Q}_{R'}^{ A,f}, \Sigma'_f\big[ Q^{f}_{R}\big];\tilde{Q}_{R'}^{B,i},\Sigma'_i\big[ Q_{R}^i\big]\!\bigg].
\end{align}
On the r.h.s.\ we have highlighted that $\Sigma'_{i,f}$ depend on what configuration $Q_R^{i,f}$ are fixed to.

When does $K$ encode dynamics -- and in which relational spacetime? The answer is subtle and depends on the pair $\Sigma_{i,f}$; see \cite{AFHMtime} for details. Our proposed sufficient condition is that whenever both $\Sigma_{i,f}$ reside in the \emph{same} integrable Cauchy foliation of $\ospace_R$, Eq.~\eqref{eqn:path-integralkernel0} encodes a dynamics in $R$-spacetime (and analogously for $R'$). 
By integrable foliation, we mean one that is classically associated with a (field-independent) timelike vector field $\rho\in\mathfrak{X}(\ospace_R)$  and Hamiltonian charge $H_{R}[\rho]$ -- the \emph{relational Hamiltonian} (general covariance fails in $\ospace_R$, so bulk Hamiltonians can exist). Infinitesimally, evolution reads $\{\hat O^A_{R},H_{R}[\rho]\}=\mathcal{L}_\rho \hat O^A_{R}$, which constitutes a $R$-reorientation. 
$K=K_{R}$ then encodes the corresponding quantum dynamics, assumed to obey standard unitary gluing rules for at least an interval of time $\tau$ labeling the slices via $\mathcal{L}_\rho\tau=1$. We interpret it as the matrix elements of the evolution operator associated with $H_{R}$, writing $K=K_{R}=\bra{Q^{A,f}_R}U_{R}(\tau_f,\tau_i)\ket{Q^{B,i}_R}$, i.e.\ holding $\Sigma_{i,f}$ fixed and letting $Q^{i,f}_{R}$ run.

While unitary, the same $K_{R}$ will \emph{not} appear as an evolution kernel in $R'$-spacetime when translated via Eq.~\eqref{eq:kerneltranslation}. First, the bounding Cauchy slices $\Sigma'_i[Q^i_R],\Sigma'_f[Q^f_R]$ differ for different matrix elements, and generically do not share an integrable foliation \cite{AFHMtime}, cf.~Fig.~\ref{fig:sharp_vs_fuzzy}. Second, evolution in $\ospace_R$ amounts to a reorientation of $R$: in an independent $R'$-perspective, there exist coordinates  $\hat O^{A'}_{R'}$ such that $K_{R}$ vanishes, unless  $Q^{A'}_{R'}$ is constant between $\Sigma'_i[Q^i_R]$ and $\Sigma'_f[Q^f_R]$, except for $Q^R_{R'}$.  By contrast, evolution kernels $K_{R'}$  correspond to fixed $\Sigma'_{i,f}$ and $R'$-reorientations that change more of the relational data in $\ospace_{R'}$. In general,  Hamiltonians for distinct frames are independent, $H_{R}\neq H_{R'}$.

Consider \emph{mixed kernels}
$K_{R\to R'}\Big[Q^{A',f'}_{R'},\Sigma'_{f'};Q^{B,i}_R,\Sigma_i\Big]$ next,
where initial and final slice are fixed in distinct relational spacetimes, and we run over the pertinent relational data on both. These are not evolution kernels in either of $\ospace_R,\ospace_{R'}$ because $\Sigma_{f'}\big[Q^{f'}_{R'}\big], \Sigma_i$ generically do not share an integrable foliation, and similarly for $R\leftrightarrow R'$. Instead, they are our proposal for \emph{frame change kernels}, 
and required to be unitary. As a sanity check, we show in \cite{AFHMtime} that in relational dynamics models with one constraint, $K_{R\to R'}$ coincides with the matrix elements of the frame change map $\mathcal{R}_{R'}(\tau'_{f'})\circ\mathcal{R}^\dag_R(\tau_i)$ of the perspective-neutral approach to QRFs (where $\mathcal{R}_R(\tau_i)$ is the Page-Wootters reduction map) \cite{delaHamette:2021oex,Hoehn:2023ehz,Hoehn:2019fsy,AliAhmad:2021adn,DeVuyst:2024pop,DeVuyst:2024uvd,delaHamette:2026rtv,Araujo-Regado:2025ejs,Giacomini:2021gei,Chen:2026kui,Castro-Ruiz:2019nnl,Apadula:2022pxk}. 

Assuming the base measure $\mathcal{D}\mu_R[\hat O_{R}]$ induces an invariant Cauchy datum measure $\mathcal{D}m_R[Q_R]$, in which $K_{R}, K_{R\to R'}$ are integrable against suitable wave functionals $\Psi\big[Q^{i,f}_{R}\big]$ (and analogously for $R'$), one finds a sequence of identities,
\begin{equation}
    \braket{\Phi_R^f}{\Psi_R^i} = \braket{\Phi^f_R}{\Psi^f_R}=\braket{\Phi^{i'}_{R'}}{\Psi^{i'}_{R'}}=\braket{\Phi^{f'}_{R'}}{\Psi^{i'}_{R'}}\,,\nonumber
\end{equation}
where states in the two perspectives are interpolated via $K_{R\to R'}$, see Fig.~\ref{fig:sequence}. Equating transition amplitudes and inner products in distinct frame perspectives, this recovers a perspective-neutral inner product \cite{AFHMtime}. It further implies an identity which enables one to link \emph{distinct} relational dynamics    
by `bending' the evolution kernel from one relational spacetime over to another, see Fig.~\ref{fig:bending}. 
\begin{figure}[h!]
    \centering
    \includegraphics[width=\linewidth]{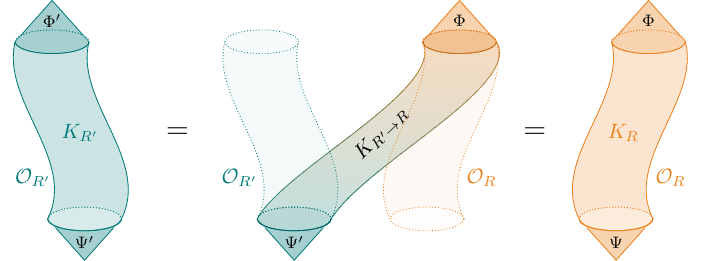}
    \caption{$K_{R\to R'}$ links transition amplitudes and inner products in distinct frame perspectives \cite{AFHMtime}.}
    \label{fig:sequence}
\end{figure}
\begin{figure}[h!]
    \centering
    \includegraphics[width=\linewidth]{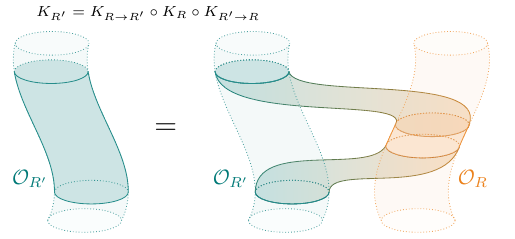}
    \caption{`Bending' the evolution kernel from $R'$- to $R$-spacetime links \emph{distinct} relational time evolutions \cite{AFHMtime}. Thus, frame change and evolution kernels are compatible.}
    \label{fig:bending}
\end{figure}

What appears as a temporally local (or `sharp') time evolution in one relational spacetime may correspond to a superposition of time evolutions in the other. Consider three slices $\Sigma'_{i_1},\Sigma'_{i_2},\Sigma'_{f'}\subset\ospace_{R'}$ inside an integrable foliation. Let $\Psi[Q_{R'}]$ be some wave function in $R'$-perspective, and let us use it as input state on both $\Sigma'_{i_1},\Sigma'_{i_2}$ for frame change and evolution kernels. The identity in Fig.~\ref{fig:bending} entails that the two unnormalized states
\begin{align}
    \Psi^{f'}_{R'}\big[Q^{f'}_{R'}\big]\! &= \!\!\sum_{j=1,2}\!\int\!\mathcal{D}m_{R'} K_{R'}\big[Q^{f'}_{R'},\Sigma'_{f'}; Q_{R'},\Sigma'_{i_j}\big]\Psi[Q_{R'}]\nonumber\\
    \!\!\!\Psi^f_{R}\big[Q^f_R\big]\!&=\!\!\sum_{j=1,2}\!\int\!\mathcal{D}m_{R'} K_{R'\to R}\big[Q^f_R,\Sigma_f;Q_{R'},\Sigma'_{i_j}\big]\Psi[Q_{R'}]\nonumber
\end{align}
 are frame change images of one another. The latter is a superposition of states on $\Sigma_f\subset\ospace_R$ obeying a sharp time evolution when concatenated with $K_R$. The former is a state on $\Sigma'_{f'}\subset\ospace_{R'}$ given by a superposition of distinct evolution kernels $K_{R'}$ acting on the same input state, which may be viewed as a superposition of time evolutions \cite{Aharonov:1990zz}. We expound in \cite{AFHMtime} how this constitutes a gravity incarnation of the QRF-dependence of temporal locality previously observed in toy models \cite{Hoehn:2019fsy,Baumann:2026lje,Castro-Ruiz:2019nnl,Giacomini:2021gei}.

\paragraph*{Relational vacua.}
The absence of a gravitational Hamiltonian in the bulk of spacetime $\spacetime$ was one of the core motivations for the no-boundary proposal \cite{Hartle:1983ai,Lehners:2023yrj,Maldacena:2024uhs}. Summing over all (inequivalent) compact Euclidean geometries to the past of a closed Cauchy surface $\Sigma\subset\spacetime$ (see \cite{Anikeeva:2026zfq} for an AdS extension) yields the Hartle-Hawking (HH) state for the data on $\Sigma$, viewed as a `vacuum without Hamiltonian'. 

The relational evolution kernels $K_R$ offer a new spectrum of distinct\,--\,albeit frame-dependent\,--\,choices for vacua in gravity. First, while the standard HH state remains a valid choice, one may impose the analogous prescription on a Cauchy slice $\Sigma_R\subset\ospace_R$, producing the \emph{relational HH state} associated with frame $R$:
\begin{equation}\label{eqn:relationalhh}
    \Psi^{\text{HH}}_{R}[Q^A_R]\equiv
    \int_{\mathcal{C}_{R}^{-}[Q^A_R]}
    \mathcal{D}\mu_R[\hat{O}_R]\,
    e^{-S_R[\hat{O}_R]} \,.
\end{equation}
Quantities on the right are analytically continued, and $\mathcal{C}_{R}^{-}[Q^A_R]$ denotes the space of compact Euclidean relational histories to the past of $\Sigma_R$ compatible with the Cauchy data $Q^A_R$ (including dressed matter fields and the dressed induced $3$-metric). Note that $\Psi^{\rm HH}_R$ coincides with $K_R$ adapted to $\mathcal{C}^-_R$. 

Now $\Sigma,\Sigma_R$ and $\Sigma_R,\Sigma'_{R'}$ are related by field-dependent maps $R[\phi]$ and $\hat U[\hat\phi]_{R\to R'}$, respectively. Thus, a fixed $\Sigma_R$ translates into fuzzy $\Sigma,\Sigma'_{R'}$, so the relational HH state is distinct from the standard one and depends on $R$.

Second, relational bulk Hamiltonians are available. We may extend the relational evolution kernel $K_R$ in $\ospace_R$ into its infinite Euclidean past, assuming $\tau$ extends that far (as it may perturbatively or in certain dS scenarios, etc.). Under standard assumptions, discussed in \cite{AFHMtime}, 
\begin{equation}\label{eqn:groundstates}
    \Psi^{\rm GS}_R\big[Q_R\big]=\int\mathcal{D}m_R[Q_R'] K_R^{\rm eucl}\big[Q_R,\Sigma_f;Q_R',-\infty\big]\Psi\big[Q'_R\big]\nonumber
\end{equation}
constitutes a (possibly distributional) state that we interpret as living in the ground sector of the relational Hamiltonian $H_R$, and which is distinct from the relational HH state. Crucially, since relational Hamiltonians $H_R,H_{R'}$ associated with different frames are in general \emph{not} unitarily related by the frame change map, neither will their ground sectors. A ground state for $R$ will typically be `seen' as excited by $R'$, in analogy to the Unruh effect and in line with previous observations of QRF-dependent vacua in non-gravitational contexts \cite{Hoehn:2023axh,Hoehn:2025pmx,Hoehn:2023ehz,cavityQED}. 

The coexistence of several notions of vacua 
reflects the different physical inputs entering their definition, so which is the appropriate one depends on the situation. Asymptotic relational ground states may be suitable for scattering scenarios, whereas a no-boundary prescription is 
appropriate when only data on a 
finite-time surface is available as in cosmology 
\cite{Hartle:1983ai,Lehners:2023yrj,Maldacena:2024uhs}.

\paragraph*{Discussion.}
We have proposed a gauge-invariant ghost-free relational path integral for non-perturbative quantum gravity, which also comes in an equivalent Faddeev-Popov gauge-fixed form. A key feature is that QRF covariance is directly built in: it is a perspective-neutral path integral that encompasses all internal QRF perspectives and transforms between them via novel frame change kernels. This new framework reveals the QRF-relativity of quantum spacetime structure: events, correlators, and time evolutions that are sharp in one perspective are generally fuzzy in another. It also gives rise to 
new relational no-boundary  and ground states. 

By coupling sources directly to relational observables, our proposal yields manifestly gauge-invariant, yet QRF-dependent generating functionals and effective actions. This prepares the ground for formulating the functional renormalization group \cite{Wetterich:1992yh, Reuter:1996cp,Thiemann:2024vjx,Ferrero:2024rvi} in relational spacetime, with the regulator tied to the physical resolution of the 
chosen frame, and to study the scale dependence of relational amplitudes and effective actions \cite{AFHM1,AFHM2}. 

A next step is to test our proposal in explicit constructions, e.g.\ in mechanical \cite{AFHMS,Benproject}, perturbative \cite{Brunetti:2013maa,Chen:2026kui}, controlled \cite{Freidel:2026stu}, or low-dimensional gravity models. The latter would permit extensions to sums over topologies.

\section*{Acknowledgments}
\begin{acknowledgments}
We would like to thank Kevin Falls, Laurent Freidel, Muxin Han, Bernard Kay, Josh Kirklin, Ben Knepper, Kasia Rejzner, Jakob Steck, Thomas Thiemann, and Tom Wetzstein for inspiring discussions that helped this project. 
SEAG, PAH and LM are supported by the Okinawa Institute of Science and Technology Graduate University.
This work was made possible through the support of the WOST, WithOut SpaceTime project (\hyperlink{https://withoutspacetime.org}{https://withoutspacetime.org}), supported by Grant ID\# 63683 from the John Templeton Foundation (JTF), and ID\#62312 grant from the JTF, as part of the ‘The Quantum Information Structure of Spacetime’ Project (QISS), as well as Grant ID\# 62423 from the JTF. The opinions expressed in this work are those of the author(s) and do not necessarily reflect the views of the John Templeton Foundation.
\end{acknowledgments}

\subsection*{Note added}
During completion of this work, \cite{Francois:2026qjv} appeared on the arXiv, which also discusses relational path integrals using a fiber bundle formulation in field space. However, our formulation may differ in significant ways, as explained in App.~\ref{sapp:contrast dressed path integral}. We also comment on the compatibility and advantages of our approach with respect to the gauge-fixed path integral for dressed fields of \cite{Falls:2025tid} in App.~\ref{sapp:dressed field functional integral}. A crucial further distinction lies in all the physical applications and qualitative predictions obtained using our framework, which have not previously been discussed in the literature.

\appendix

\section{Notation}\label{app:notation}
In this appendix we collect the main notation in this manuscript in order of appearence. 

\begin{itemize}[noitemsep,
    leftmargin=1.7cm,
    labelwidth=1.6cm,
    labelsep=0.1cm,
    align=left]

\item[{$\mathcal{F}$}] Field space.
\item[{$\mathcal{M}$, $g$}] Spacetime manifold and metric respectively.
\item[{$\mathcal{G}$}] Gauge symmetry group; corresponding to small diffeomorphisms in the main text.
\item[{$\hat{\mathcal{F}}$}] Base space, $\hat{F}:=\mathcal{F}/\mathcal{G}$.
\item[{$\mathfrak{G}$}] Field space metric.
\item[{$R[\phi]$}] Dynamical frame field.
\item[{$\mathcal{U}_R$}] Field-dependent spacetime domain of a frame field $R[\phi]$.
\item[{$\mathcal{U}_{\mathcal{O}}$}] Image of the frame (field-dependent coordinate system).
\item[{$\mathcal{O}_R$}] Relational spacetime associated with $R$.
\item[{$f$}] Small diffeomorphism, $f\in\mathcal{G}$.
\item[{$T$}] Gauge-covariant local quantity, such as a tensor field.
\item[{$O_{T|R}$}] Relational observable associated with $R$. In particular, $O_{g\mid R}$ is a metric on $\mathcal{O}_R$.
\item[{$o$}] Point in $\mathcal{O}_R$.
\item[{$O^A_R$}] Relational observable coordinatizing $R$-horizontal directions.
\item[$R^\alpha$] Field-space coordinates of frame fields $R[\phi]$.
\item[$\hat{O}^A_R$] Base space relational coordinates.
\item[{$U_{R\rightarrow R'}[\phi]$}] Gauge-invariant spacetime coordinate transformation, see \eqref{eq:frame-change-map}.
\item[$\hat{\epsilon}$] Volume form on $\physf$.
\item[{$\mathcal{D}\mu[\hat O_R]$}] Measure of the relational path integral, see \eqref{eq:basespacemeasure}.
\item[{$\hat{O}_R$}] Base space relational observable.
\item[{$\gamma_{\alpha\beta}$}] Gauge orbit metric.
\item[{$\phi$}, $\hat{\phi}$] Field configuration in $\mathcal{F}$ and $\physf$, respectively.
\item[{$S_R$}] Relational classical action, obtained by pushfoward to $\mathcal{O}_R$.
\item[{$Z$}]  Relational path integral, see \eqref{eq:relational-path-integral}.
\item[{$\mu_R[\hat{O}_R]$}] Gauge-invariant field space measure.
\item[{$\varphi^i$}] Components of the dynamical fields residing in the action, with $i$ a DeWitt index encoding spacetime and internal indices, as well as $x\in\spacetime$.
\item[$\gf{R}$] Fixed frame configuration defining gauge-fixing function $\chi_{R,\gf{R}}[\phi]$.
\item[$\chi_{R,\gf{R}}$] Gauge-fixing function, $\chi_{R,\gf{R}}=R-\gf{R}$.
\item[{${\Delta}_{R}$}] Faddeev–Popov operator.
\item[$\hat{O}_{\tilde{T},R'}$] Scalarized base space relational observable relative to $R'$, see \eqref{eq:relobsrewrite}.
\item[$\hat{U}_{R\to R'}$] Frame change map for base space relational observables.
\item[{$J_R$}] Background source field in $\mathcal{O}_R$.
\item[{$Z_R[J_R]$}] Relational generating functional, see  \eqref{eq:relational-generating-functional}. 
\item[{$W_R[J_R]$}] Relational Schwinger functional, see  \eqref{eq:relational-generating-functional}.
\item[{$\Gamma_R[{O}_R]$}] Relational effective action, see \eqref{eq:relational-effective-action}.
\item[{$\bar{O}$}] Mean field value of observable $O$.
\item[{$\Sigma_R$}] Cauchy slice in $\mathcal{O}_R$. $R$-label is often dropped when no ambiguity can arise.
\item[{$Q_R^{A,n}$}] Defined as $Q_R^{A,n}:=\hat{O}_R^A[\hat{\phi}]|_{\Sigma_n}$, $n=i,f$. Relational field configurations on field-independent initial and final Cauchy slices $\Sigma_i$ and $\Sigma_f$ respectively in $\mathcal{O}_R$.
\item[{$K$}] Path integral evolution kernel for relational dynamics in $\ospace_R$.
\item[${\physf}^{fi}$] Base space restricted to histories compatible with data specified on field-independent initial and final Cauchy slices $\Sigma_i$ and $\Sigma_f$ respectively in $\mathcal{O}_R$.
\item[$\tilde{Q}^{A,n}_{R'}$] Scalarized observables relative to $R'$ on field-independent initial and final Cauchy slices $\Sigma_i$ and $\Sigma_f$ respectively in $\mathcal{O}_R$.
\item[{$\mathfrak{X}$}] Space of vector fields.
\item[{$\rho_R$}] Timelike vector field in $\mathfrak{X}(\mathcal{O}_R)$.
\item[$\mathcal{L}_\rho$] Lie-derivative along $\rho\in\mathfrak{X}(\mathcal{O}_R)$.
\item[{$H_R[\rho]$}] Relational bulk Hamiltonian and associated with $\rho\in\mathfrak{X}(\ospace_R)$.
\item[$K_R$] Evolution kernel encoding relational dynamics in $R$-perspective.
\item[{$U_R(\tau^i,\tau^f)$}] Unitary evolution operator between initial and final times $\tau_{i,f}$ specified by Cauchy slices $\Sigma_{i,f}$ in a foliation of $\mathcal{O}_R$.
\item[{$K_{R\to R'}$}] Frame change kernel.
\item[{$\mathcal{D}m_R[Q_R]$}] Invariant Cauchy datum measure induced from $\mathcal{D}\mu_R[\hat{O}_R]$.
\item[$\Psi_R^{\text{HH}}$] Relational Hartle-Hawking state associated with frame $R$, see \eqref{eqn:relationalhh}.
\item[$\Psi_R^{\rm {GS}}$] State belonging to the ground sector of the relational Hamiltonian $H_R$, see \eqref{eqn:groundstates}.
\item[$\delta$] Field space exterior derivative.
\item[$\diff$] Spacetime differential, commuting with $\delta$.
\item[{$\iota_X \omega$}] Interior product, i.e.\ contraction of $X\in\mathfrak{X}(\fspace)$ into a form $\omega$ on $\fspace$.
\item[{$\mathsf{L}_{X}$}] Lie derivative with respect to $X\in\mathfrak{X}(\fspace)$.
\item[{$\pi$}] Map $\mathcal{F}\rightarrow\physf$.
\item[$\omega^\alpha_\parallel$] DeWitt field space connection induced by $\fmetric$ and $\gamma$.
\item[{$\fmetric_H$, $\fmetric_V$}] Respectively horizontal and vertical components of $\fmetric$ with respect to the DeWitt connection.
\item[{$\varpi_R$}] Generalized Maurer-Cartan form on $\mathcal{F}$, see  \eqref{connection}.
\item[$\sigma_{\gf{R}},\mathcal{X}_{\gf{R}}$] Section defined by the gauge-fixing condition $R=\gf{R}$, and its image $\mathcal{X}_{\gf{R}}=\mathrm{Im}(\sigma_{\gf{R}})$.
\item[$\hat{\fmetric}$] Base space metric, $\fmetric_H=\pi^\star\hat{\fmetric}$ determining the volume form $\hat{\epsilon}$.
\item[{$\mathcal{A}_\xi$}] Anomaly of the path integral measure.
\item[$\mathfrak{U}_R$] Domain of $\mathbf{R}(\phi,x)=R[\varphi](x)$ in $\fspace\times\spacetime$.
\item[$\mathcal{V}$, $\hat{\mathcal{V}}$] Neighborhoods on $\fspace$ and $\physf$, respectively.
\item[$\mathcal{A}_a$, $\mathscr{A}$] Spacetime atlas and field space meta-atlas, respectively.
\item[$\approx$] On-shell identity.
\item[$\delta_\lambda$] Variation of frame-dependent functional along a one-parameter family of frames $\{R_\lambda\}_{\lambda\in\mathbb{R}}$.
\item[$\nabla_\lambda$] Generalized derivative along a one-parameter family of frames $\{R_\lambda\}_{\lambda\in\mathbb{R}}$, see \eqref{eqn:generalizeddlambda}.
\end{itemize}

\section{Dynamical frames for the diffeomorphism group and relational observables}\label{app_frames}

Let us review the key features of dynamical frames for gravity, as introduced in \cite{Goeller:2022rsx,Carrozza:2022xut} and used throughout our work, by comparing them to tetrads of the Lorentz group, i.e.\ to arguably the paradigmatic example of internal frames in physics. Observing that dynamical frames for gravity are to the diffeomorphism group what tetrads are to the Lorentz group may help the uninitiated reader  better appreciate the meaning of the \emph{physical} frame covariance they give rise to in generally covariant theories. We note that quantum reference frames (QRFs) in mechanical models admit the same comparison with tetrads \cite{Hoehn:2023ehz,delaHamette:2021oex}. There is thus a universal underlying covariance structure.

We begin with tetrads in special relativity.\footnote{The argument can be easily extended to the full Poincar\'e group by supplementing the internal frame (i.e.\ the tetrad) with a position degree of freedom.} Acting on Minkowski space vectors (extended to general tensors in the obvious manner), we may view them as a map
\begin{equation}
    e^\mu_a:\spacetime\to\ospace
\end{equation}
with distinct in- and output spaces, where $\mu$ denotes the spacetime and $a$ the frame index. $\spacetime$ is Minkowski space, while $\ospace\simeq\spacetime$ denotes what one may call \emph{relational Minkowski space}. It constitutes the home for how the tetrad `sees' spacetime. Thanks to $\eta_{\mu\nu}\,e^\mu_a\,e^\nu_b=\eta_{ab}$, we have that $e$'s configurations\,--\,henceforth called \emph{orientations}\,--\,take value in the Lorentz group, i.e.\ ${e^\mu_a\in\rm{O}(3,1)}$. Coming with two distinct indices, the tetrad does not constitute a representation of the Lorentz group by itself, rather it is a `Lorentz transformation from the input to the output space'.

We saw in the main body that dynamical frame fields in gravity are nothing but field-dependent coordinate systems,
\begin{equation}
    R[\phi]:\spacetime\supset\mathcal{U}_R[\phi]\to\mathcal{U}_\ospace\subset\ospace_R\,,
\end{equation}
thus constituting maps with distinct in- and output spaces. The input is spacetime $\spacetime$, while $\ospace_R$ is relational spacetime. When $R[\phi]$ is a diffeomorphism (as it is in many examples), we have that $R[\phi]\in\rm{Diff}(\spacetime,\ospace_R)$, i.e.\ the configurations of the frame field $R$\,--\,called its global \emph{orientations}\,--\,take value in the diffeomorphisms from the input to the output space, akin to the tetrad. As a field, it also takes local values, or orientations, $o\in\ospace_R$ at $x\in\spacetime$, which is why $\ospace_R$ was also called `local orientation space' in \cite{Goeller:2022rsx}.

Let us now turn to external frame transformations. In special relativity, these are the Lorentz transformations $\Lambda^\mu{}_\nu\in\rm{O}(3,1)$ acting on the spacetime index. As such, they act on the input space $\spacetime$ and thereby on \emph{all} fields; the tetrad transforms covariantly $e^\mu_a\mapsto\Lambda^\mu{}_\nu e^\nu_a$ under these. 

In gravity, the external frame transformations are given by spacetime diffeomorphisms and these include the gauge transformations. Indeed, these amount to changes of \emph{external} (i.e.\ field-independent) coordinate frames. Spacetime diffeomorphisms thus act on the input space of the frame $R[\phi]$ and they act on \emph{all} fields in $\spacetime$. As emphasized in the main body, dynamical frames in gravity are \emph{gauge-covariant} under small diffeomorphisms,
\begin{equation}
    R[f_\star\phi]=R[\phi]\circ f^{-1}\,,\qquad f\in\mathcal{G}\subset\rm{Diff}(\spacetime)\,,
\end{equation}
transforming as a set of $\dim\spacetime$-scalars (constructed out of the available field content of the theory).

A tetrad also admits \emph{reorientations}. These are given by the Lorentz transformations $\Lambda_a{}^b\in\rm{O}(3,1)$ acting on the frame index, i.e.\ on the output space given by relational spacetime $\ospace$. As such, they act exclusively on the frame and trivially on all other fields. They change the relation between other degrees of freedom and the tetrad (see below). Clearly, external frame transformations $\Lambda^\mu{}_\nu$ and internal frame reorientations $\Lambda_a{}^b$ constitute distinct representations of the Lorentz group and so commute. Hence, reorientations are purely internal and independent of any external Lorentz frame.

Returning to gravity, \emph{reorientations} are given by field-independent diffeomorphisms in the output space of the frame, i.e.\ in relational spacetime
\begin{equation}\label{eq:reorientations}
    R[\tilde{f}_\star\phi]=\tilde{f}\circ R[\phi]\,,\qquad\tilde{f}\in{\rm Diff}(\ospace_R)\,,
\end{equation}
acting on the `other side' of the frame. As such, they leave all other (independent, see App.~\ref{app:geometry}) fields invariant. Reorientations commute with spacetime diffeomorphisms and hence are gauge-invariant \emph{physical} transformations. It is important to stress that by reorientations we mean \emph{active} diffeomorphisms in $\ospace_R$ that act on all \emph{dynamical} fields in relational spacetime, but not on background fields (more on this below in App.~\ref{app_relfibercoord}). In other words, reorientations change the field configuration, in contrast to passive redefinitions of $R$ by a field-independent diffeomorphism in $\ospace_R$ at a fixed point $\phi\in\fspace$ in field history space. 

The distinction is important because a theory that is generally covariant in $\spacetime$ will typically not be generally covariant in relational spacetime $\ospace_R$ where the frame will lead to background structure \cite{AFHMtime}. The theory can thus distinguish the two. Reorientations are important: boundary symmetries admit the interpretation of reorientations of an edge mode frame \cite{Carrozza:2022xut,Carrozza:2021gju,Araujo-Regado:2024dpr,Araujo-Regado:2025ejs}, while relational time evolution amounts to a reorientation of the `clock' \cite{AFHMtime,DeVuyst:2024pop,DeVuyst:2024uvd}. Reorientations are also the cause for type conversions in gravitational von Neumann algebras \cite{DeVuyst:2024uvd}.

Next, how does an internal frame `see' the other fields? Consider some tensor $T^{\mu\cdots}{}_{\nu\cdots}$ in $\spacetime$ in special relativity. The corresponding description relative to the tetrad $e$ is given by the frame-dressed observable
\begin{equation}
    T^{a\cdots}{}_{b\cdots}\coloneqq e^a_\mu\,e^\nu_b\cdots T^{\mu\cdots}{}_{\nu\cdots}\,.\label{eq:Tab}
\end{equation}
By contracting all spacetime indices, clearly $T^{a\cdots}{}_{b\cdots}$ is invariant under external Lorentz frame transformations. It is the same tensor `pushed forward' to relational spacetime $\ospace$ and its components correspond to how an observer in a lab associated with the tetrad would measure this tensorial quantity. Thus, it is a frame-relational observable. Note, in particular, that $T^{a\cdots}{}_{b\cdots}$ changes under reorientations $\Lambda_a{}^b$.

In fact, we can slightly tweak the description by sending it back to Minkowski space $\spacetime$ via a \emph{fixed} configuration $\tilde{e}^\mu_a\in{\rm O}(3,1)$ of the same frame, while keeping $e^\mu_a$ unconstrained:
\begin{equation}
    T^{\alpha\cdots}_{\tilde{e}}{}_{\beta\cdots}\coloneqq (\tilde{e}_a^\alpha \,e^a_\mu)\,(\tilde{e}^b_\beta\,e^\nu_b)\cdots T^{\mu\cdots}{}_{\nu\cdots}\,.\label{eq:Talphabeta}
\end{equation}
When $\tilde{e}$ is a fixed `background field' which neither transforms under `active' external frame changes or reorientations, we have that $T_{\tilde{e}}^{\alpha\cdots}{}_{\beta\cdots}$ is independent of external frames and equivalent to Eq.~\eqref{eq:Tab}. The new form in Eq.~\eqref{eq:Talphabeta} makes it clear that $T_{\tilde{e}}^{\alpha\cdots}{}_{\beta\cdots}$ is the relational observable that encodes the value of the bare tensor components $T^{\mu\cdots}{}_{\nu\cdots}$ when tetrad $e$ is in orientation $\tilde e$. Indeed, since Eq.~\eqref{eq:Talphabeta} is independent of external frames, we can `gauge fix' $e=\tilde{e}$ without changing the values of $T_{\tilde{e}}^{\alpha\cdots}{}_{\beta\cdots}$ and in this `gauge' we have $T_{\tilde{e}}^{\alpha\cdots}{}_{\beta\cdots}\equiv T^{\mu\cdots}{}_{\nu\cdots}$.

Back in gravity, the procedure is analogous. Suppose $T[\phi]$ is some small diffeomorphism gauge-covariant dynamical quantity in $\spacetime$, i.e.\ obeying $T[f_\star\phi]=f_\star T[\phi]$, $f\in\mathcal{G}$, such as a tensor field (see \cite{Carrozza:2022xut,Baldazzi:2021fye} for a generalization beyond gauge-covariant quantities). The corresponding description relative to frame field $R[\phi]$ is the frame-dressed observable 
\begin{equation}
    O_{T|R}[\phi]\coloneqq\left(R[\phi]\right)_\star T[\phi]\,,
\end{equation}
which is nothing but the push forward of $T$ to relational spacetime $\ospace_R$ by the frame itself. That is, $O_{T|R}$ is now a tensor field on $\ospace_R$ if $T$ was on $\spacetime$. It measures the components of $T$ at the dynamically defined event $x$ where $R[\phi](x)=o\in\ospace_R$. As shown in \cite{Goeller:2022rsx}, this encompasses and unifies previous approaches to dressed and relational observables in gravity.

Clearly, by gauge covariance,
\begin{align}
    O_{T|R}[f_\star\phi]&=\left(R[f_\star\phi]\right)T[f_\star\phi] = \left(R[\phi]\circ f^{-1}\right)_\star f_\star T[\phi]\nonumber\\
    &=\left(R[\phi]\right)_\star T[\phi]=O_{T|R}[\phi]\,,
\end{align}
so $O_{T|R}$ is (small) spacetime diffeomorphism-invariant. It is, however, not invariant under reorientations, i.e.\ relational spacetime diffeomorphisms. Indeed, by Eq.~\eqref{eq:reorientations}, we have
\begin{equation}
    O_{T|R}[\tilde{f}_\star\phi]=\tilde{f}_\star O_{T|R}[\phi]\,,\qquad\tilde{f}\in{\rm Diff}(\ospace_R)\,.
\end{equation}

In analogy to Eq.~\eqref{eq:Talphabeta}, we can also implement a tweak, \begin{equation}
    O^{\tilde{R}}_{T\mid R}[\phi] \coloneqq (\tilde{R}^{-1}\circ R[{\phi}])_\star T[{\phi}]\,,\label{eq:newrelobs}
\end{equation}
where $\tilde{R}$ is any \emph{fixed} (so field-independent) global orientation of $R$. So $\tilde{R}^{-1}\circ R[\phi]\in{\rm Diff}(\spacetime)$ is a field-dependent \emph{spacetime} diffeomorphism and since $\tilde{R}$ is a background field, $O^{\tilde{R}}_{T|R}$ remains small diffeomorphism-invariant and equivalent to $O_{T|R}$. This makes clear that $O^{\tilde{R}}_{T|R}$ is a \emph{relational observable}: gauge fixing\footnote{It is a gauge-covariant field, so we can gauge fix it.} $R[\phi]\overset{!}{=}\tilde{R}$ clarifies that $O_{T|R}^{\tilde{R}}$ measures the components of $T$ conditional on frame $R$ being in orientation $\tilde{R}$. Indeed, in that gauge we have $O^{\tilde{R}}_{T|R}\equiv T$. In other words, the components of the gauge-invariant relational tensor $O^{\tilde{R}}_{T|R}$ \emph{coincide} with the components of the non-invariant bare tensor $T$ in the field configuration $\phi$, where $R[\phi]=\tilde{R}$.

Finally, let us consider frame changes, beginning once more with special relativity. 
Suppose we have a second tetrad $e'^{\mu}_{a'}$ with its own frame index $a'$ and relational spacetime $\ospace'$ and wish to transform the tensor $T$ from $e$- to $e'$-perspective. Invoking the normalization $e^\mu_a\,e^a_\nu=\delta^\mu_\nu$ of the tetrads, we then have
\begin{align}
    T^{a'\cdots}{}_{b'\cdots}= e'^{a'}_{\mu}\,e'^\nu_{b'}\cdots T^{\mu\cdots}{}_{\nu\cdots} = \Lambda^{a'}{}_{a}\,\Lambda_{b'}{}^{b}\cdots T^{a\cdots}{}_{b\cdots}\,,\label{eq:lorentz}
\end{align}
where
\begin{equation}
    \Lambda^{a'}{}_{a}\coloneqq e'^{a'}_{\mu}\,e^\mu_a\in{\rm O}(3,1)\label{eq:lorentz2}
\end{equation}
is the Lorentz transformation from $e$- to $e'$-perspective, i.e.\ from $\ospace$ to $\ospace'$, given by the relational observable describing $e'$ relative to $e$. Note that Eq.~\eqref{eq:lorentz} constitutes a change of frame dressing, i.e.\ a change of relational observable.

\begin{figure}
    \centering
    \includegraphics[width=\linewidth]{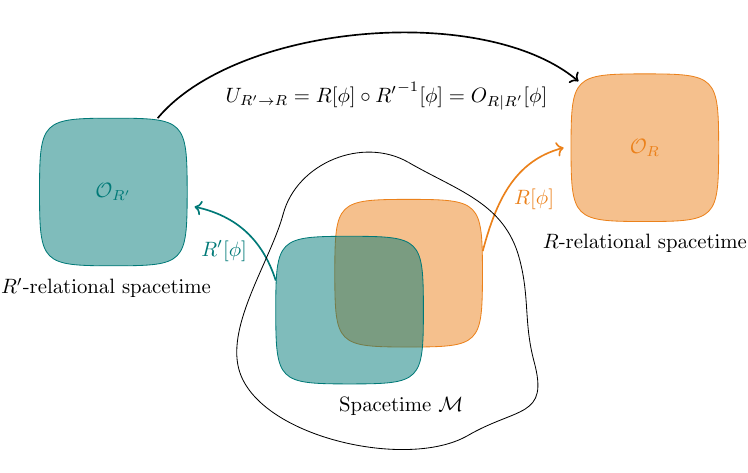}
    \caption{Changes of dynamical frame in gravity are field-dependent coordinate transformations from one relational spacetime to another and given by the relational observable $O_{R|R'}[\phi]$  describing  the new frame relative to the old one.}
    \label{fig:frame_app_fig}
\end{figure}

Next, we return to gravity. Since dynamical frame fields are nothing but field-dependent coordinate systems, it is clear that frame changes are nothing but field-dependent coordinate transformations, see Fig.~\ref{fig:frame_app_fig}. Indeed, suppose we have two distinct frame fields $R,R'$. The analog of Eq.~\eqref{eq:lorentz2} is the small gauge-invariant coordinate change
\begin{align}
    U_{R\to R'}[\phi]&\coloneqq (R'[\phi])\circ(R[\phi])^{-1}\nonumber\\& = (R[\phi])_\star R'[\phi] = O_{R'|R}[\phi]\,,\label{eq:framechangeapp}
\end{align}
where the second equality follows from $R'$ transforming as scalars. When $R,R'$ are both diffeomorphisms, $U_{R\to R'}$ is a gauge-invariant diffeomorphism from relational spacetime $\ospace_R$ to relational spacetime $\ospace_{R'}$ and given by the relational observable describing $R'$ relative to $R$. Similarly, the analog of Eq.~\eqref{eq:lorentz} simply reads
\begin{equation}
    O_{T|R'}[\phi] = \left(U_{R\to R'}[\phi]\right)_\star\,O_{T|R}[\phi]\,.
\end{equation}

Invoking the tweaked version of relational observables in Eq.~\eqref{eq:newrelobs}, we instead have the field-dependent frame change
\begin{align}
\tilde{U}_{R\to R'}[\phi]&\coloneqq \left(\tilde{R}'^{-1}\circ R'[\phi]\right)\circ\left(\tilde{R}^{-1}\circ R[\phi]\right)^{-1}\nonumber\\
&=\tilde{R}'^{-1}\circ O_{R'|R}[\phi]\circ\tilde{R}\in{\rm Diff}(\spacetime)\,,\label{eq:tweakedchange}
\end{align}
given by a spacetime diffeomorphism. The tweaked relational observables then simply transform as 
\begin{equation}
    O^{\tilde R'}_{T|R'}[\phi] = \left(\tilde{U}_{R\to R'}[\phi]\right)_\star\,O^{\tilde R}_{T|R}[\phi]\,.
\end{equation}

In conclusion, tetrads in special relativity and dynamical frame fields in gravity obey completely analogous covariance structures. Tetrads are to the Lorentz group what dynamical frame fields are to the (small) diffeomorphism group.

\section{Geometric setup in field space}
\label{app:geometry}

In this appendix, we review basics of the field space bundle formulation, exhibit a few new results regarding field space connections, and provide further technical details on the construction of the relational path integral \eqref{eq:relational-path-integral}. In Fig.~\ref{fig:comparizon_field} we depict key features of the field space bundle, which includes a summary of the notation in this appendix.
\begin{figure}
    \centering
    \includegraphics[height=\linewidth]{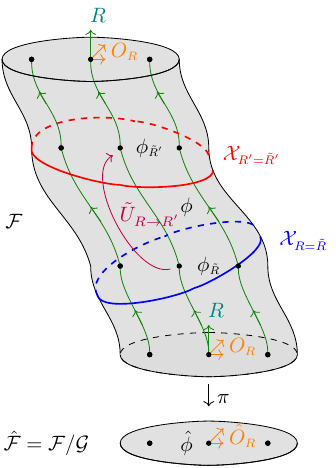}
    \caption{
    Pictorial summary of the field space bundle $\fspace$ underlying our construction of the relational path integral. Green curves depict the fibers, i.e.\ gauge orbits of the small diffeomorphism gauge group $\mathcal{G}$. Field configurations $\phi\in\fspace$ correspond to the entire history of the dynamical fields on spacetime $\spacetime$. Field configurations in the base space of orbits $\physf$ are gauge equivalence classes of histories (entire fibers) and denoted by $\hat\phi$.\\
    Each atlas of frame fields (cf.~App.~\ref{app_metaatlas}) defines a family of local sections $\mathcal{X}_{\tilde{R}}$ (red and blue circles) in $\fspace$, determined by fixing the frame fields $R$ into orientation $\tilde R$. Each such atlas also determines a horizontal connection, defining vertical (tangential to the fibers) and horizontal (tangential to the sections) splits. This can be exploited to build a relational \emph{fiber-adapted coordinate system} $(R^\alpha,O^A_R)$ on patches of $\fspace$ from each atlas (cf.~Apps.~\ref{app_relfibercoord} and~\ref{app_metaatlas}). Its associated relational observables parametrize horizontal spaces, while $R$ itself parametrizes the fibers. The relational observables $O^A_R$ are $\pi$-pullbacks of observables $\hat O^A_R$ on $\physf$.\\
    The change of frames in Eq.~\eqref{eq:tweakedchange} is a spacetime diffeomorphism $\tilde{U}_{R\to R'}[O_{R'|R}]$ depending on the relation of the frame fields. Acting on the bundle it amounts to nothing else than changing local section ${\mathcal{X}_{\tilde R}\to\mathcal{X}_{\tilde R'}}$.  
}
    \label{fig:comparizon_field}
\end{figure}

\subsection{Basic field space geometry}

The set of permissible kinematical field configurations {$\phi$ on some spacetime manifold $\mathcal{M}$} will be denoted by $\fspace$ and referred to as \emph{field space}. {$\fspace$ may encode suitable boundary conditions, but no equations of motion.} In the following, we treat $\fspace$ as an infinite-dimensional manifold, locally coordinatized by the field components $\varphi^i(x)$. We employ DeWitt notation, whereby spacetime and internal indices are combined into a single label, $\varphi^i(x)\equiv \varphi^i$, with $i$ understood to encode both the discrete field indices and the dependence on $\mathcal{M}$.

On $\fspace$ we introduce the exterior derivative $\delta$, satisfying $\delta^2=0$ 
and commuting with the spacetime differential $\diff$, cf.~\cite{anderson1992introduction}. Acting on functionals $F:\fspace\to\mathbb{R}$, it is given by ${\delta F = \frac{\delta F}{\delta \varphi^i}\,\delta \varphi^i\equiv F_{,i}\delta\varphi^i}$, where repeated indices denote both a sum over internal labels and an integration over $\spacetime$. In this notation, vector fields on $\fspace$, $X\in \vfields$, where $\vfields$ denotes the space of vector fields on $\mathcal{F}$, take the form $X = X^i\,{\delta}/{\delta \varphi^i}$, with $\iota_X\delta\varphi^i=X^i$, $\iota$ representing contraction.

Spacetime diffeomorphisms act on $\fspace$ by pushforward, $\phi\to f_\star\phi$.
In this work, we focus on {arbitrary generally covariant theories whose Lagrangians obey ${L[f_\star\phi,\eta]=f_\star L[\phi,\eta]}$, $\forall f\in\mathrm{Diff}(\spacetime)$, where $\eta$ denotes any background fields \cite{Goeller:2022rsx,Giulini:2006yg,Anderson1971-ANDCIA}. Gauge transformations are constituted by `small diffeomorphisms' $\mathcal{G}\subset\mathrm{Diff}(\spacetime)$,\footnote{{Internal matter gauge symmetries can be treated independently \cite{Donnelly:2016auv,Gomes:2018dxs,Carrozza:2021gju,Araujo-Regado:2024dpr,Kunstatter:1991kw}; we assume these have already been accounted for.}} often forming a subgroup, cf.~\cite{Goeller:2022rsx}}. Infinitesimally, their action is generated by a Lie algebra {$\mathrm{Lie}(\fggroup)\subset \mathfrak{X}(\spacetime)$, and encoded in $\fspace$ by} vector fields $\fundvfields:\mathrm{Lie}(\fggroup)\to \vfields$, such that $\fundvfields:\xi\mapsto \fundvfields_\xi$ {with $\
\iota_{X_\xi}\delta\varphi^i=\mathcal{L}_\xi\varphi^i$}. We will{ write} $X_\alpha=X(T_\alpha)$, where $T_\alpha$ is a basis of $\mathrm{Lie}(\fggroup)${, so that} infinitesimal gauge-transformations of a functional $F$ read
\begin{equation}
    \delta_\xi F\equiv X_\xi(F){=\iota_{X_\xi}\delta F}=\xi^\alpha X_\alpha^iF_{,i}{=\mathsf{L}_{X_\xi}F}\,,
\end{equation}
{where $\mathsf{L}_{X_\xi}$ denotes the field space Lie derivative.

Henceforth, we assume that the \emph{gauge orbits},\footnote{Note, however, that in gravitational and typical matter gauge theories, gauge symmetries as \emph{redundancies} (in the sense of degenerate directions of the presymplectic structure) usually arise only on the space of classical solutions, e.g.\ see \cite{Lee:1990nz,Harlow:2019yfa,Carrozza:2021gju,Carrozza:2022xut}.} to which the $X_\alpha$ are tangent, give $\fspace$ the structure of a principal fiber bundle with \emph{typical fiber} $\mathcal{G}$ and base space $\physf\coloneqq\fspace/\mathcal{G}$, referred to as the \emph{space of orbits}, which encodes the gauge-invariant information. The bundle projection is denoted $\pi:\fspace\to\physf$. Since the action $S=\int_\spacetime L$ obeys $\delta_\xi S=0$, it descends to a functional on $\physf$.}

{Following DeWitt's proposal \cite{DeWitt:2003pm,DeWitt:1967ub,DeWitt:1967yk,DeWitt:1980jv,Parker:2009uva,Kunstatter:1991ds,DeWitt:1998eq, gospel}}, we invoke a gauge-invariant metric $\fmetric$ on $\fspace$, ${\mathsf{L}}_{\fundvfields_\xi}\fmetric=0$, usually constructed from the matrix of second derivatives of the kinetic term of the action (and gauge-breaking terms). For example, in pure gravity this leads to the well-known DeWitt metric \cite{DeWitt:2003pm}. We then define the gauge orbit metric
\begin{equation}\label{eq:gaugeorbitmetric}
    \gamma_{\alpha\beta}\equiv \fmetric(\fundvfields_\alpha,\fundvfields_\beta),
\end{equation}
assumed to be invertible.  $\fmetric$ defines a connection $\omega^\alpha_\parallel=\fmetric(\cdot,X_\beta)\gamma^{\alpha\beta}$ {\cite{DeWitt:2003pm,DeWitt:1998eq}, locally splitting $T_\phi\fspace=\mathcal{V}_\phi\oplus\mathcal{H}_\phi$ into vertical (the gauge orbits) and \emph{orthogonal} horizontal subspaces, $\mathcal{H}=\ker\omega_\parallel$, associated with the} projector ${P^i_j=\delta^i_j-X^i_\alpha(\omega^\alpha_{\parallel})_j}$. {This induces a local block decomposition of} $\fmetric$ into horizontal and vertical components $\fmetric=\fmetric_H+\fmetric_V$, where $\fmetric_V=\gamma_{\alpha\beta}\omega^\alpha_\parallel\omega^\beta_\parallel$, and $\fmetric_H=[\fmetric_\perp]_{ij}\omega_\perp^i\omega_\perp^j$, with $\omega_\perp^i=P^i_j\delta\varphi^j$ and $[\fmetric_\perp]_{ij}\equiv P^m_i P^n_j\fmetric_{mn}$. 

DeWitt proposed to introduce fiber-adapted coordinates on (patches of) $\fspace$ \cite{DeWitt:2003pm,DeWitt:1967ub,DeWitt:1967yk,DeWitt:1980jv,Parker:2009uva}, splitting gauge-invariant coordinates from ones that parametrize the fibers, to go along with the metric block diagonalization. As discussed in the main body, DeWitt \cite{DeWitt:2003pm} (and others later \cite{Pawlowski:2003sk,Donkin:2012ud}) eventually abandoned this idea, considering such coordinate systems as largely symbolical without immediate physical interpretation. In particular, DeWitt \cite{DeWitt:2003pm} worried about the fact that the gauge-invariant coordinates could only be defined implicitly and that they may depend non-locally on the $\varphi^i$ (in $\spacetime$). This, in turn, would render unclear how to impose initial and final data using such variables, as well as how to define a notion of advanced propagators which underlie his construction of the field space measure. 

In this work, we revisit and address this challenge using the notion of dynamical reference frames \cite{Goeller:2022rsx,Carrozza:2022xut,Freidel:2025ous}. The key step is to realize that one must now formulate everything in \emph{relational} spacetime $\ospace_R$, i.e.\ spacetime `as seen' by frame $R$ (see App.~\ref{app_frames}), as opposed to the standard spacetime $\spacetime$ one began with. Indeed, relational observables, while highly non-local in $\spacetime$, are local in $\ospace_R$. Furthermore, they can be made explicit and have concrete physical interpretations as describing some quantity relative to the chosen frame. As shown in \cite{Goeller:2022rsx,Marolf:2015jha}, they obey microcausality in $\ospace_R$ with respect to its causal structure (defined via the dressed metric $O_{g|R}$) at least for frames locally constructed from the fundamental fields in the action. Thus, in relational spacetime one may well define the notion of advanced propagators. Similarly, one may impose initial and final data in terms of relational observables on (possibly only patches of) Cauchy slices in relational spacetime in the usual sense. This is what we exploit in the main body when defining evolution and frame change kernels, which we discuss in greater detail together with relational Hamiltonians in \cite{AFHMtime}. Therefore, dynamical frames permit us to address the challenges noted by DeWitt in relational terms. In other words, the price to pay for solving these challenges is that the resolutions are frame-dependent (yet can be translated into other frame perspectives). However, in a generally covariant theory without absolute structures this must be expected.

\subsection{Relational fiber-adapted coordinates}\label{app_relfibercoord}

{Let us now construct relational fiber-adapted coordinates on $\fspace$ invoking the dynamical frames for generally covariant theories reviewed in App.~\ref{app_frames}.

Infinitesimally, the frame fields' components $R^\alpha$ transform as
\begin{equation}\label{eqn:framevariation}
   \delta_{\xi} R^\alpha=\xi^\beta {\Delta_R}_{\beta}^\alpha  \,,\qquad {\Delta_R}^\alpha_\beta\equiv X^i_\beta R^\alpha_{,i}\,,
\end{equation}
{and gauge covariance implies that the \emph{Faddeev-Popov operator} $\Delta_R$ is locally invertible. Hence,} the set of frame labels $\alpha$ is in one-to-one correspondence with a basis $T_\alpha$ of $\mathrm{Lie}(\fggroup)$, allowing us to employ the same Greek indices, $\alpha,\beta,\ldots$.

In regions where $\Delta_R$ in \eqref{eqn:framevariation} is invertible, $R^\alpha$ parametrizes the fibers, and so gauge fixing $R$ to any \emph{fixed} configuration, $R^\alpha[\phi]=\gf{R}^\alpha$, defines a local section ${\sigma_{\gf{R}}:\hat{\mathcal{V}}\subset\physf\to\fspace}$ of the projection 
${\pi:\fspace\to\physf}$, establishing a one-to-one correspondence between  
  ${\mathcal{X}_{\gf{R}}\equiv \sigma_{\gf{R}}(\hat{
  \mathcal{V}})}$ 
and  $\hat{\mathcal{V}}$. Denote by $\{\varphi^A[\varphi^i]\}$ any independent set of  functions, such that the $\varphi^A|_{\mathcal{X}_{\gf{R}}}$ parametrize
$\mathcal{X}_{\gf{R}}$. Whenever the $\varphi^i$ are  tensor fields, $\varphi^A$ may be chosen gauge-covariant. Clearly, the restriction ${O^A_R|_{\mathcal{X}_{\gf{R}}}=(\gf{R})_\star\varphi^A|_{\mathcal{X}_{\gf{R}}}}$ of the relational observables ${O^A_R\coloneqq O_{\varphi^A|R}}$ thus provides an equivalent parametrization of $\mathcal{X}_{\gf{R}}$. However, since the $O^A_R$ are gauge-invariant, they \emph{also} descend to coordinates $\hat O^A_R$ on $\hat{\mathcal{V}}$ via ${O^A_R=\hat O^A_R\circ\pi}$. Thus, relational observables  are complete on patches of the base. 
In summary, 
$(R^\alpha,O^A_R)$ locally separate vertical and horizontal directions of $\fspace$, thus constituting a \emph{relational fiber-adapted coordinate system}, with $R^\alpha$ parametrizing the fibers and $O^A_R$ the base. This establishes our claim from the main text.

Therefore, each choice of frame provides us with a horizontal connection {given by a (generalized) Maurer-Cartan form on $\fspace$ \cite{Gomes:2016mwl,Freidel:2021dxw,Carrozza:2022xut,Donnelly:2016auv,Ciambelli:2022vot}}
\begin{equation}\label{connection}
    \varpi_R{\coloneqq-\delta R^{-1}\circ R}= \Delta_R^{-1}\delta R=\omega_\parallel+\Delta_R^{-1} R_{,i}\omega_\perp^i\,,
\end{equation}
satisfying $\iota_{X_\xi}\varpi_R=\xi$ and
defining a frame-dependent decomposition of $T_{{\phi}}\mathcal{F}=\mathcal{V}_{R{,\phi}}\oplus {\mathcal{H}_{R,\phi}}$, with ${\mathcal{H}_{R,\phi}}\equiv \ker\varpi_R$ {(see \cite[App.~A]{Carrozza:2022xut} for intuition behind $\varpi_R$). The last two equalities in Eq.~\eqref{connection} link $\varpi_R$ with the Faddeev-Popov operator and DeWitt's connection $\omega_\parallel$, and will be proven in \cite{AFHM2}. $\mathcal{H}_{R,\phi}$ is tangent to the $\mathcal{X}_{\gf{R}}$ and spanned by the $\delta/\delta O^A_R$}.

Let us now return to the notion of frame changes and reorientations reviewed in App.~\ref{app_frames} and consider their effect in the field bundle $\fspace$.
First, the frame changes in Eq.~\eqref{eq:framechangeapp} change the description
 from one frame perspective to another, i.e.\ from one relational spacetime to another, but not the field configuration. This transformation induces changes of fiber-adapted coordinates and, via $\varpi_R\to\varpi_{R'}$, of horizontal distribution at a fixed $\phi\in\fspace$.

Second, the \emph{frame reorientations} \cite{Carrozza:2022xut,Goeller:2022rsx,Freidel:2025ous} in Eq.~\eqref{eq:reorientations} change the configuration of a fixed frame field $R[\phi]$ without changing the remaining fields. They are active relational spacetime diffeomorphisms infinitesimally generated by  $Y_{\rho_R}\in\mathfrak{X}(\fspace)$, corresponding to (field-independent) $\rho_R\in\mathfrak{X}(\mathcal{O}_R)$, and defined by $\iota_{Y_{\rho_R}}\delta R^{-1}\equiv\mathcal{L}_{\rho_R} R^{-1}$ and ${\iota_{Y_{\rho_R}}\delta O^A_R=\mathcal{L}_{\rho_R} O^A_R}$.
As these transformations change relational observables, they are \emph{physical} transformations, having a horizontal component for any connection, $\pi_\star Y_{\rho_R}\neq0$; e.g.\ boundary symmetries admit the interpretation of reorientations \cite{Carrozza:2022xut,Carrozza:2021gju,Araujo-Regado:2024dpr,Araujo-Regado:2025ejs}, as does time evolution, see main body and \cite{AFHMtime}. We have $\iota_{Y_{\rho_R}}\varpi_R=-(R[\phi])^\star\rho_R$ \cite{Carrozza:2022xut}, so reorientations have a vertical component and shift the horizontal section within the family $\{\sigma_{\gf{R}}\}_{\gf{R}}$. Furthermore, $\iota_{Y_{\rho_R}}\varpi_{R'}=0$ for any  independent frame $R'[\varphi^A]$, therefore $R$'s reorientations are entirely horizontal for $R'$.

\subsection{Field space measure}\label{sapp:field measure}

 As a next step toward constructing a relational path integral, we write DeWitt's $\fspace$-measure \cite{DeWitt:2003pm,DeWitt:1998eq,Parker:2009uva,Bonanno:2025xdg} in relational coordinates as key ingredient of our path integral. Using the block-diagonal decomposition of $\mathfrak{G}$ and  $\delta O^A_R={O_R^A}_{,i}\omega_\perp^i$, we have}
\begin{equation}\label{eqn:horizontalrepar}
\fmetric_H=[\invfmetric_R]_{AB}\delta O_R^A\delta O_R^B\,,\quad [\invfmetric_R]_{AB}=[\fmetric_\perp]_{ij}\varphi^i_{,A}\varphi^j_{,B}\,.
\end{equation}
Using standard path integral notation, we write formally $\mathcal{D}\varphi=\bigcurlywedge_i\delta\varphi^i$, $\mathcal{D}O_R=\bigcurlywedge_A\delta O^A_R$, $\mathcal{D}g=\bigcurlywedge_\alpha\omega^\alpha_\parallel$ (the fiber volume form), and $\mathcal{D}R=\bigcurlywedge_\alpha \delta R^\alpha$, where $\curlywedge$ denotes the wedge product on $\fspace$, and, suppressing $\curlywedge$ between terms, obtain the volume form as
\begin{align}\label{eqn:fieldpsacemeasure}
    \mathcal{D}\varphi\sqrt{\det\fmetric}
    &=
    \mathcal{D}O_R\,\mathcal{D}{g}\,
    [\det \invfmetric_R\,\det\gamma]^{1/2}\\
    &=
    \mathcal{D}O_R\,\mathcal{D}R\,
    [\det \invfmetric_R\,\det\gamma]^{1/2}\,
    (\det \Delta_R)^{-1}\,.\nonumber
\end{align}
In the second line, we used the last equality in Eq.~\eqref{connection}.

Crucially, when $\mathfrak{G}$ is gauge-invariant, then so are both $\det\gamma$ and $\det \invfmetric_{R}$ 
\cite{Kunstatter:1991kw,Parker:2009uva}, meaning that  
\begin{equation}\label{eqn:mu}
    \mu_R[O_R]\coloneq [\det \invfmetric_{R}\det\gamma]^{1/2}[O_R]
\end{equation}
is a functional of relational observables only. {Hence, it is the pullback of a measure on $\hat\fspace$: $\mu_R[O_R]=\mu_R[\hat O_R]\circ\pi$. Specifically, $\mathfrak{G}_H=\pi^\star\hat{\mathfrak{G}}$ defines a metric $\hat{\mathfrak{G}}$ with volume form $\hat{\epsilon}=\mathcal{D}\hat{O}_R[\det\hat\fmetric]^{1/2}$ on $\hat\fspace$ via pullback. The gauge-invariant part of the volume form in Eq.~\eqref{eqn:fieldpsacemeasure} is then the pullback of the base space form
\begin{equation}\label{eq:basespacemeasureapp}
\mathcal{D}\mu_R[\hat O_R]\coloneqq\mathcal{D}\hat{O}_R\,\mu[\hat O_R]=\hat{\epsilon}\times[\det\gamma]^{1/2}[\hat{O}_R]\,.
\end{equation}
$\hat\epsilon$ is invariant under arbitrary field redefinitions on $\hat\fspace$, while $\det\gamma$ transforms as a scalar because Eq.~\eqref{eq:gaugeorbitmetric} is defined independently of the choice of fiber-adapted coordinates (though it depends on the  $\mathrm{Lie}(\fggroup)$-basis). Thus, Eq.~\eqref{eq:basespacemeasureapp} is covariant under arbitrary frame changes and reorientations\,--\,a key result for frame covariance in the main body.

\subsection{Spectrum of choices for the field space measure}\label{app_manymeasures}

Let us pause to consider different strategies for choosing a field space measure and discuss their qualitative distinctions, especially as regards quantum frame covariance. The options hinge on whether or not orbit information is retained and the general difference between first quantizing then reducing a theory (`Dirac quantization'), or, conversely, first reducing then quantizing it (`reduced quantization') \cite{Kunstatter:1991ds,Kunstatter:1991qe}. Of course, in practice, any of these procedures will necessitate a regularization to properly define the measure.

Here, we followed DeWitt's geometric strategy 
\cite{DeWitt:2003pm}, starting from a gauge-invariant metric $\fmetric$ on the  full field space $\fspace$, chosen independently of any particular frame. Since 
the metric, and hence the measure, are defined before choosing a frame, the construction is manifestly perspective-neutral, yields a consistent quantum frame covariance, and retains the information encoded in the gauge orbit metric $\gamma$. This is a covariant version of first quantizing then reducing the theory.

This does not mean that the field space metric, and hence the measure, is
unique: its choice is part of the definition of the quantum theory, and different choices may lead to inequivalent quantizations. After all, quantization remains an \emph{a priori} ambiguous procedure both in canonical or path integral language. The question is therefore how such a metric should be specified. A standard proposal, due to DeWitt, is to construct it from the Hessian of the action, particularly its kinetic term
\cite{DeWitt:2003pm}. In \cite{Benproject}, we show how to do this explicitly in a mechanical toy model, recovering the canonical (Dirac-quantized) perspective-neutral formulation of it \cite{Vanrietvelde:2018pgb}.

This prescription can be implemented in two ways. First, one may apply it directly on the full field space $\fspace$. In a gauge theory, however, the Hessian has zero modes along the gauge orbits, so obtaining a non-degenerate field space metric requires the introduction of gauge-breaking terms. Since these terms are not unique, this construction inherits the usual quantization
ambiguities and ultimately only an experiment can single out the `right' prescription. The choice of gauge breaking term may or may not be related to a choice of frame. Frame-neutral choices do exist as we demonstrate in the toy model of \cite{Benproject}. This is the route we  have in mind for the gravity case. 

Alternatively (and generally inequivalently), one may first determine the horizontal metric through Eq.~\eqref{eqn:horizontalrepar}, using the Hessian of the relational action or, equivalently, the (advanced \cite{DeWitt:2003pm}) propagator on relational spacetime expressed in terms of relational observables associated with a chosen frame. The aim is then to build a full base space metric $\fmetric$ and therefrom a volume form $\hat\epsilon$ using relational data. Due to the doubly local nature of frames, in order to cover all of $\physf$ and $\spacetime$, this would require a meta-atlas of frames discussed in App.~\ref{app_metaatlas}, i.e.\ to `patch up' $\hat\epsilon$. Up to here, this procedure is consistent with reduced quantization
 discussed below. A global field space metric is then
reconstructed by supplementing this horizontal metric with a choice of metric, or measure, along the gauge orbits, which, if non-trivial, renders it distinct from reduced quantization \cite{Kunstatter:1991ds,Kunstatter:1991qe}. In this formulation, part of the measure
ambiguity is tied to the  meta-atlas of frames used to define the horizontal sector, the other to the choice of orbit metric. Different choices may again lead to inequivalent quantum theories, but each completed
full field space measure may still be viewed as defining  a perspective-neutral theory: the (meta-atlases of) frames label the quantization prescription, rather than a preferred perspective
within the resulting theory. Unitary frame covariance would then exist within a choice of such perspective-neutral theory, but not across such choices.

This situation has an analog in canonical Dirac quantization. Classically, the constraints may be replaced by any locally invertible, possibly phase-space-dependent recombination, without changing the constraint
surface or the gauge orbits; the change can be absorbed into a corresponding phase-space-dependent redefinition of lapse and shift. This equivalence need
not survive quantization, and different representatives of the same classical constraint system can yield unitarily inequivalent physical Hilbert spaces
(see \cite{Kuchar:1991qf} for a general discussion and \cite{Gielen:2020abd,Gielen:2021igw} for concrete examples). 

This ambiguity can again be related to a choice of frame: different representations of the constraints naturally single out candidate frame variables canonically conjugate, at least
locally, to the chosen constraints. This determines the notion of gauge covariance of the QRF; indeed, QRFs are typically modeled using positive operator-valued measures that are covariant with respect to the unitary representation of the gauge group \cite{delaHamette:2021oex,Hoehn:2019fsy,Marchetti:2024nnk} (see also \cite{Carette:2023wpz,Fewster:2024pur,Garmier:2025soc} for physical symmetry groups).  Conversely, selecting a frame induces an adapted constraint representation via a preferred phase-space-dependent choice of lapse and shift. The resulting Dirac-quantized theories may therefore be unitarily inequivalent perspective-neutral theories, while each  still contains all frame perspectives and one may change unitarily between them. However, one cannot unitarily change frame across theories.

Reduced quantization,\footnote{Here, by reduced quantization we mean its covariant implementation. A canonical reduced phase space quantization may differ in an important respect in gravity: since generic spacetime diffeomorphisms are not projectable to a fixed reduced phase space \cite{Lee:1990nz,Pons:1996av,Pons:2003uc}, the induced measure need not remain
diffeomorphism-invariant and may develop an anomaly \cite{Han:2009bb}. Such an anomaly may in principle be removed by appropriately dressing the measure, following, e.g., the procedure of \cite{Francois:2026qjv}; see below. As argued in \cite{Francois:2026qjv}, however, this transfers the anomaly to frame changes.}  
on the other hand, can be understood as corresponding to a
different choice of \emph{class} of measures \cite{Kunstatter:1991ds,Kunstatter:1991qe} (see also \cite{Engle:2009ba,Han:2009bb,Vanrietvelde:2018dit,Thiemann:2024vjx,Ferrero:2024rvi}). Reduced quantization corresponds to retaining only the horizontal measure $\hat\epsilon$ described above, \emph{without} supplementing it by a measure $\det\gamma$ along the gauge orbits. $\hat\epsilon$ may be constructed using a choice of meta-atlas, as discussed above. The resulting measure is therefore insensitive to the gauge fibers and is manifestly tied to the chosen meta-atlas of frames. Specifically, in contrast to first quantizing then reducing above, where one treats all degrees of freedom on an essentially equal footing, it is hard to see how one may devise a method for selecting an $\physf$-measure that does not depend on a choice of frames.  Nevertheless, as long as $\hat\epsilon$ defines a volume form on $\physf$ frame covariance can again hold; the arguments of the main body would in this case still apply, yielding a frame-neutral base space path integral $Z'$, formally differing from Eq.~\eqref{eq:relational-path-integral} by the $\det\gamma$ contribution (and the precise choice of $\hat\epsilon$). Thus, unitary frame covariance in terms of our passive transformations within a fixed theory (from a distinct class of theories than above) could still hold in this case (see also \cite{Thiemann:2026vaj}). 

However, as with distinct perspective-neutral theories above, there will be no unitary change of frames \emph{across} reduced quantized theories, meaning theories defined by distinct base measures $\hat\epsilon$ which may explicitly depend on the choice of meta-atlas of frames. Presumably, the frame change anomaly reported in \cite{Francois:2026qjv} when changing the frame-dependent dressed measure, introduced to eliminate possible gauge anomalies, refers to this unitary inequivalence across theories; see App.~\ref{sapp:contrast dressed path integral} for details. This aligns with previous observations in a canonical setting that in general one cannot perform unitary QRF changes across different reduced quantizations of the same theory, where one first reduces the classical theory into a given frame perspective and then quantizes the result \cite{Vanrietvelde:2018dit}.

In summary, there is a spectrum of admissible measures, whose ambiguities are
part of the definition of the quantum theory and  lead to inequivalent quantizations. Several of these (especially reduced methods) are inherently tied to a choice of (meta-atlas of) frames. 
By contrast, the  first-quantize-then-reduce strategy adopted  in this work identifies, in principle, a route towards a class of outright frame-neutral measures that is non-anomalous under both gauge transformations and frame changes. Although this does not eliminate the intrinsic measure ambiguity, it ensures compatibility with both gauge invariance and perspective neutrality. As will be shown in \cite{Benproject}, this route can be explicitly implemented in mechanical toy models. The next step is to demonstrate that it also works in concrete field-theoretic models.

\subsection{Relational path integral and possible anomalies}

Now, we have all the elements to support the path integral construction of the main text. 

Given the volume form  in Eq.~\eqref{eqn:fieldpsacemeasure}, we may  formally define a gauge-invariant path integral, assuming henceforth that $\fmetric$ is gauge-invariant.  Integrating a gauge-invariant functional $F[O_R]$ over $\fspace$, the fiber contribution, $\int\mathcal{D}g$, produces the divergent volume of $\mathcal{G}$. Dropping this redundant factor yields a non-anomalous relational path integral that is formally finite and, using Eq.~\eqref{eq:basespacemeasure}, may be equivalently written as a base space integral
\cite{AFHM1,AFHM2}:
\begin{equation}
    Z
    =
    \int_{\hat\fspace}\hat{\epsilon}\,\sqrt{\det\gamma[\hat\phi]}\,e^{iS[\hat\phi]}=
    \int_{\hat\fspace} \mathcal{D}
    \mu[\hat{O}_R]e^{iS_R[\hat{O}_R]}\,,
    \label{eq:relational-path-integral-appendix}
\end{equation}
where hatted variables denote base space integration variables and $\hat\phi\in\hat\fspace$. In the second equality, we used that $S[\varphi]=\int_\spacetime L=\int_{\ospace_R}R_\star L=:S_R[O_{A|R}]$ when the frame field $R$ is spacetime global, i.e.\ when ${U}_\spacetime=\spacetime$. Our construction is also written using a single frame label for $\fspace$. In general, a frame field $R$ is neither global in $\spacetime$ and $\fspace$, and so we need to `patch up' using a meta-atlas of frames on both $\spacetime$ and $\fspace$, see App.~\ref{app_metaatlas}.

As in the operator quantization of gravitational null rays in \cite{Freidel:2025ous,Freidel:2026stu}, the relational description thus formally leads to a non-anomalous quantum theory, and it would be interesting to explore connections.

Let us briefly entertain the possible source of  anomalies if one changed steps in the construction. For non-invariant $\fmetric$,} ${\mathsf{L}}_{X_\xi}\mu_R\neq 0${;}  the measure is {then} said {to be} \emph{anomalous} and {non-}projectable {to} $\hat\fspace$. {Supposing that $\fmetric_H$ is invariant,\footnote{Otherwise, it cannot be the pullback $
\fmetric_H=\pi^\star\hat\fmetric$ of a base metric.    

} we can isolate} the anomaly $\mathcal{A}_\xi$ by ${\mathsf{L}}_{X_\xi}\mu_R\equiv \mathcal{A}_\xi\,\mu_R$, {finding} $2\mathcal{A}_\xi={\mathsf{L}}_{X_\xi}\log\det\gamma$. Any anomaly in the measure {then} originates from the vertical 
volume element associated with the gauge orbits.

Lastly, while $Z$ is manifestly gauge-invariant, we can build an equivalent gauge-fixed full field space version of it.  Before descending to the  base integral, we can fix the frame configuration via $\chi_{R,\gf{R}}[\varphi]=R[\varphi]-\gf{R}\overset{!}{=}0$, obtaining the equivalent Faddeev-Popov gauge-fixed $\fspace$-integral 
\begin{align}\label{eq:Z_relational2}
Z&=\int\mathcal{D}{R}\,\delta[\chi_{R,\gf{R}}[{R}]]\int\mathcal{D}{O}_R\,\mu_R[{O}_R]\,e^{iS_R[{O}_R]}\nonumber\\
    &=\int_\fspace\mathcal{D}{\varphi}\sqrt{\det\fmetric}\,\delta[\chi_{R,\gf{R}}[\varphi]]\det \Delta_{R}\, e^{iS[\varphi]}\,,
\end{align}
where in the second line we used Eq.~\eqref{eqn:fieldpsacemeasure}. The conceptual differences between gauge-invariant and gauge-fixed form are discussed in the main text.

\section{Relational meta-atlas and `patching up' spacetime and path integrals}\label{app_metaatlas}

As noted in several places, frame fields $R[\phi]$ are in general defined only locally in both spacetime $\spacetime$ \emph{and} field space $\fspace$. They constitute gauge-covariant coordinate systems on spacetime and we use them to build (small) diffeomorphism-invariant descriptions by pushing spacetime fields forward to the relational spacetimes. Due to the doubly local nature of the frames, we need many such relational spacetimes to have a consistent global description encompassing all of $\spacetime$ and $\fspace$. 

In particular, the construction of our relational path integral, generating functionals and effective actions invoke integrals over both field space and relational spacetime. We must thus explain what it means to carry out such integrals when the relational coordinate systems that we use are defined locally in $\fspace$ and $\spacetime$, i.e.\ how to `patch up' such integrals. This is, of course, nothing special to our setup and applies to any integrals over manifolds ranging over multiple coordinate patches and is standardly addressed by invoking `partitions of the identity' on the pertinent manifold, e.g.\ see \cite{vitagliano2024primer}. In this appendix, we explain how to adapt this tool from the integration theory on manifolds to our setting. This construction invokes and significantly refines the meta-atlas of dynamical frames to cover both $\spacetime$ and $\fspace$ initially discussed in \cite{Goeller:2022rsx}.

Let us now begin by building a relational atlas for $\spacetime$ in a neighborhood of field space. To encode the field-dependence of a frame field ${R[\phi]:\mathcal{U}_R[\phi]\to \mathcal{O}_R}$ and its spacetime dependence, we introduce a map ${\mathbf{R}:\mathfrak{U}_R\to\mathcal{O}_R}$, where $\mathfrak{U}_R\subset\fspace\times\spacetime$, such that ${\mathbf{R}(\phi,x)=R[\phi](x)}$. The spacetime domain of the frame is then 
    \begin{equation}
        \mathcal{U}_R[\phi]=\{x\in\spacetime\mid (\phi,x)\in\mathfrak{U}_R\}\,. 
    \end{equation}
Let us now consider a `gauge-saturated' patch on field space $\mathcal{V}_a\subset\fspace$, namely a patch containing complete gauge orbits, so that $\mathcal{V}_a=\pi^{-1}(\hat{\mathcal{V}}_a)$ for some $\hat{\mathcal{V}}_a\subset\physf$.  We assume that a collection of such patches, labeled by $a$, covers $\fspace$:  $\cup_a\mathcal{V}_a=\fspace$. Let us further assume that 
    \begin{equation}
        \bigcup_{i\in I_a}\mathcal{U}_{ai}[\phi]=\spacetime\,,\qquad\forall\phi\in\mathcal{V}_a\,,
    \end{equation}
    where $I_a$ is an index set for spacetime frame fields $R_{ai}[\phi]$ (i.e.\ gauge-covariant coordinate systems) covering $\spacetime$ on $\mathcal{V}_a$.
    Then $\mathcal{A}_a=\{\mathcal{U}_{ai},R_{ai}\}_{i\in I_a}$ is a relational spacetime atlas for all $\phi\in\mathcal{V}_a$. Whenever $\mathcal{U}_{ai}[\phi]\cap\mathcal{U}_{aj}[\phi]\neq \emptyset$, transition maps for the above atlas are given, for all $\phi\in\mathcal{V}_a$, by the frame change map $U^a_{i\to j}[\phi]$ defined in Eq.\ \eqref{eq:frame-change-map}. Thus, $U^a_{i\to j}[\phi]$ represents a field-dependent spacetime coordinate change, interpolating between the descriptions of the same kinematical quantities relative to the frames $R_i[\phi]$ and $R_j[\phi]$.
    
    As a shorthand, denote by $O_{ai}$ the relational coordinates $\{O^A_{R_{ai}}\}$ for each frame field $R_{ai}$. For each $\mathcal{V}_a$, $R_a\equiv \{R_{ai}\}_{i\in I_a}$ and\footnote{Note that \eqref{eqn:defoa} retains the  local representatives $O_{ai}$ in each overlap of the domain of the pertinent frames. But the compatibility condition ensures that they are not regarded as independent, thereby consistently accounting for the apparent redundancy.} \begin{equation}\label{eqn:defoa}
        O_{a}\equiv \left\{(O_{ai})_{i\in I_a}\mid O_{ai}=(U^a_{j\to i})_\star O_{aj}~\text{on every overlap}\right\},
    \end{equation}
   define a fiber-adapted coordinate system $\chi_a\equiv (R_a, O_a)$ for the patch $\mathcal{V}_a$. Thus, $\mathscr{A}\equiv \{\mathcal{V}_a,\chi_a\}_{a}$ is a meta-atlas, i.e., an atlas built from frame fields for both $\fspace$ \emph{and} $\spacetime$.

   By projection, one has the following atlas on $\physf$: ${\hat{\mathscr{A}}=\{\hat{\mathcal{V}}_a,\hat{O}_a\}_{a}}$, where $\hat{O}_a$ and $\hat{\mathcal{V}}_a$ are the descendant of $O_a$ and $\pi$-projection of $\mathcal{V}_a$, respectively. On a non-trivial field space overlap $\mathcal{V}_a\cap\mathcal{V}_b\neq \emptyset$, one must compare the two relational atlases $\mathcal{A}_a$ and $\mathcal{A}_b$, which can be done per point $\phi$ in the intersection via the frame change map ${U_{ai\to bj}[\phi]=R_{bj}\circ R_{ai}^{-1}}$. Collectively, these induce field-space transition maps $T_{a\to b}$ between fiber-adapted coordinates and, again by projection, transitions maps on the base space, $\hat{T}_{a\to b}$. In particular, these transition maps define the base space coordinate changes $\hat{O}_{R'}^{A'}[\hat{O}_R^A]$ discussed in the main text, which underlie the frame covariance of base-space functionals.
  To summarize, the meta-atlas $\mathscr{A}$ incorporates  a collection of spacetime atlases $\mathcal{A}_a$ associated with a collection of patches covering field space and accounts for regions of overlap between the frames in both $\mathcal{M}$ and $\hat{\mathcal{F}}$.

Having defined a relational atlas on $\spacetime$ and the associated $\physf$ meta-atlas, we can express integrals on $\spacetime$ and $\physf$ in terms of the corresponding local relational coordinates, using partitions of unity subordinate to the respective covers. For spacetime integrals, such partitions of unity can be given by functions $\eta_{ai}:\mathcal{V}_a\times\spacetime\to [0,1]$, such that
    \begin{equation}\label{eq:identity res}
        \mathrm{supp}(\eta_{ai})\subset\mathfrak{U}_{ai}\,,\qquad\sum_{i\in I_a}\eta_{ai}(\phi,x)=1_a\,,
    \end{equation}
    for each patch $a$. A particularly important example of such integrals is given by the action $S[\phi]$, for which we can write
     \begin{align} \label{eq:action}  S[\phi]&=\sum_{i\in I_a}\int_{\mathcal{U}_{ai}[\phi]}\eta_{ai}[\phi](x)\,L[\phi](x)\nonumber\\
        &=\sum_{i\in I_a}\int_{\mathcal{O}_{ai}}\hat{\eta}_{ai}[\hat{O}_{ai}](o_{ai})\,L_{ai}[\hat{O}_{ai}](o_{ai})\nonumber\\
        &\equiv S_{a}[\hat{O}_a]\,,
    \end{align}
    where $\hat{\eta}_{ai}[\hat{O}_{ai}]=(R_{ai}[\phi])_\star\eta_{ai}[\phi]$ and $L_{ai}[\hat{O}_{ai}]=(R_{ai}[\phi])_\star L[\phi]$ are the dressed versions of $\eta_{ai}$ and $L$, respectively.\footnote{Note that $\hat{\eta}_{ai}$ satisfies the dressed unity condition $\sum_{i\in I_a}\hat{\eta}_{ai}[\hat{O}_{ai}]=1_a$.}

Analogously, let us then define a partition of unity $\hat{\rho}_a$ on base space $\physf$ subordinate to $\hat{\mathcal{V}}_a$, with
    \begin{equation}
        \mathrm{supp}(\hat{\rho}_a)\subset\hat{\mathcal{V}}_a\,,\qquad\sum_a{\hat{\rho}}_a=1.
    \end{equation}
    $\hat{\rho}_a$ can then be used to express any base space integral in terms of local relational coordinates. For instance, the relational path integral \eqref{eq:relational-path-integral} now takes the form 
    \begin{equation}\label{eq: new patched integral}
        Z=\sum_a\int_{\hat{\mathcal{V}}_a}\mathcal{D}\mu_a[\hat{O}_a]\,\hat{\rho}_a[\hat{O}_a]\,e^{iS_a[\hat{O}_a]}\,.
    \end{equation}
  Eqs.\ \eqref{eq: new patched integral} and \eqref{eq:action} clearly show that both the relational $\spacetime$ and $\physf$ atlases play a crucial role in the definition of the path integral, as expected.

    As emphasized above, the same construction applies more generally to integrals on $\spacetime$, $\fspace$, or $\physf$. In particular, it extends to integrals over Cauchy slices, as required for the definition of relational Hamiltonians, where the local relational presymplectic currents\,--\,and hence the resulting presymplectic forms\,--\,must be consistently glued across overlapping frame patches (see \cite{AFHMtime} for some discussion).

\section{Comparison with previous relational path integral proposals}\label{app_otherintegrals}
{In this appendix, we compare the relational path integral \eqref{eq:relational-path-integral} with previous proposals for path integrals formulated in terms of dressed fields in \cite{Falls:2025tid} and \cite{Francois:2026qjv}. The reader is referred to \cite{Diaz:2026zdr,Wei:2025guh} for other interesting approaches to relational path integrals.  

We emphasize that besides the formal construction of a relational path integral, which was also studied in somewhat different form in \cite{Falls:2025tid,Francois:2026qjv}, the present work reveals entirely novel qualitative physical predictions regarding frame covariance of correlation function of relational observables, frame-dependent temporal locality of transition amplitudes, relational ground states, relational no-boundary states, and gauge-invariant generating functionals, including the relational effective action.

\subsection{Functional integral for dressed fields in \texorpdfstring{\cite{Falls:2025tid}}{}}\label{sapp:dressed field functional integral}
Here, we show that the relational path integral in \eqref{eq:relational-path-integral} reproduces the proposal in \cite{Falls:2025tid} for the functional integral of dressed fields, and in addition makes more explicit the underlying measure, which was left unspecified in \cite{Falls:2025tid}.

This can be seen by choosing a frame $\mathscr{R}[\varphi]=\tilde{R}^{-1}\circ R[\varphi]$, with $\tilde{R}$ a fixed configuration of $R$, see Eq.\ \eqref{eq:newrelobs} and App.~\ref{app_frames}. 
Upon gauge-fixing the frame configuration to $0\overset{!}{=}\chi_{\mathscr{R},\mathrm{Id}}[\varphi]=\mathscr{R}[\varphi]-\mathrm{Id}(x)=\mathscr{R}[\varphi]-x$, the path-integral \eqref{eq:Z_relational} reduces to
\begin{align}\label{eq:new path integral}
Z=\int_\fspace\mathcal{D}{\varphi}\sqrt{\det\fmetric}\,\delta[\chi_{\mathscr{R},\mathrm{Id}}[\varphi]]\, e^{iS[\varphi]}\,,
\end{align}
since $\det\Delta_{\mathscr{R}}=1$ for the above $\mathscr{R}$-gauge-fixing. The resulting expression corresponds to the functional path integral for dressed fields proposed by \cite{Falls:2025tid}. 

Even though \eqref{eq:new path integral} has been gauge-fixed, the path-integral is ghost-free, as $\det\Delta_{\mathscr{R}}=1$. Note that the above arguments can be repeated analogously for any gauge theory, by gauge fixing the (appropriately chosen) frame field to the identity of the gauge group, leading again to a ghost-free quantum theory.  Equivalently, as shown in Eq.\ \eqref{eq:Z_relational}, the Faddeev-Popov determinant arises from the change from fiber-adapted to non-fiber-adapted
coordinates. However, when the frame field is fixed to the identity, one has
\begin{equation}
    O_{T\mid \mathscr{R}}[\varphi]\vert_{\mathscr{R}=\mathrm{Id}}=(\mathscr{R}[\varphi])_\star T[\varphi]\vert_{\mathscr{R}=\mathrm{Id}}=T[\varphi]\vert_{\mathscr{R}=\mathrm{Id}}\,,
\end{equation}
so that the corresponding coordinate transformation, and hence the
Faddeev--Popov determinant, becomes trivial. Thus the relational path integral \eqref{eq:relational-path-integral} offers a natural explanation for the disappearance of ghost contributions in \cite{Falls:2025tid}. Moreover, through
Eq.~\eqref{eq:new path integral}, it also determines the appropriate
integration measure for this gauge-fixed formulation, coming from the manifestly gauge-invariant formulation.

\subsection{Dressed path integral proposal in \texorpdfstring{\cite{Francois:2026qjv}}{}}\label{sapp:contrast dressed path integral}

Although \cite{Francois:2026qjv} formally works with a dressed path integral,
its main focus is on addressing gauge anomalies through dressing-field techniques, rather than on providing a precise characterization of the dressed
path integral itself. The latter is defined somewhat ambiguously, making it difficult to identify
precisely how their construction differs from ours. As anticipated by the discussion in App.~\ref{sapp:field measure}, these ambiguities can be traced
back to different choices of measure.

For instance, their dressed path integral may be interpreted as arising from
the dressing of a gauge-anomalous, path-integral top form, whereas we start from a gauge-invariant global field space metric.  It would then remain unclear, however, how the finite integration over the base space appearing in their Eq.~(10) is obtained. This interpretation also seems difficult to reconcile with their Eq.~(12), since the cocycle property should apply to the integrand rather than to the path integral itself. Indeed, Eq.~(12) relates two quantities that are independent of the field configuration\,--\,the bare and dressed path integrals\,--\,through a
field-dependent cocycling dressing field. This tension may stem from treating the path integral somewhat oddly as a local field space form, rather than as an inherently non-local integration. The same issue is reflected in the potentially misleading notation $Z[\phi],Z[\bar\phi]$, and analogously in that used for its dressed counterpart and generating functionals.

Alternatively, one may consider the dressed path integral to be defined using an appropriate reduced measure, cf.~App.~\ref{app_manymeasures}. Several statements in \cite{Francois:2026qjv} seem to support this interpretation,  for example that gauge fixing is `unnecessary (impossible even)' \cite{Francois:2026qjv}. As explained in App.~\ref{app:geometry}, however, a perspective-neutral path integral coming from the full field space $\fspace$ can be gauge-fixed, leading to an equivalent Faddeev-Popov form. Moreover, gauge anomalies are naturally encoded in the gauge orbit volume factor $\det\gamma$ entering the measure $\mu_R$, and can therefore only be addressed at the level of the full field space path integral. By contrast, reduced quantization removes this factor already at the classical level, thereby eliminating the very structure in which such anomalies arise. If the dressed path integral of \cite{Francois:2026qjv} is indeed based on a reduced measure,
then gauge anomalies are not cancelled within that construction; rather, they are absent by construction.

Regardless of the exact definition of their dressed path integral, however, the frame anomaly arising from their `anomaly seesaw mechanism'
signals a breakdown of frame covariance, in contrast with the construction adopted here, which is based on measures that are non-anomalous under both gauge transformations and frame changes; see App.~\ref{sapp:field measure}.

A further potentially significant departure from our approach concerns their definition of the generating functionals. In \cite{Francois:2026qjv} it is not explicitly stated whether the bare sources are spacetime background fields, although the notation appears to suggest this. If so, this would constitute a major difference from our approach. As emphasized in the main text, the relational sources in Eq.~\eqref{eq:relational-generating-functional} are not obtained by dressing
spacetime background sources. Since such background fields do not transform under active diffeomorphisms, while the frame does, dressing them would inherit
the transformation properties of the frame and would therefore fail to produce gauge-invariant sources. The resulting generating functional would consequently
be gauge-dependent, unlike in our formulation.

\section{Observables and fuzzy correlators}\label{app_fuzzy}
Here, we offer further detail surrounding the discussion of fuzzy correlators in Eq.~\eqref{eq:fuzzycorrelator}, especially on the properties of the observables $\hat{O}_{\tilde{T}_i|R'}$ inside the $n$-point function in $R'$-perspective. We will clarify how to think of them as relational observables in $R'$-perspective and provide more information on how to write them as functionals of the relational coordinates $\hat O^{A'}_{R'}$. We illustrate the differences between sharp and fuzzy events in Fig.~\ref{fig:sharp_vs_fuzzy}.

Gauge-invariant observables on $\physf$ and $\fspace$ are equivalent through pullback: $O =\hat{O}\circ \pi$. As it will now be convenient to decompose the relevant relational observables into spacetime quantities and explore their transformation properties under diffeomorphisms, we will work with the unhatted (equivalent) version of the relational observables. Recall from Eq.~\eqref{eq:relobsrewrite} that
\begin{equation}\label{eq:tilderelational}
    \hat{O}_{\tilde{T}_i|R'}=\hat{O}_{{T}_i|R}\circ \hat U_{R'\to R}\,,
\end{equation}
which, using Eqs.~\eqref{eq:frame-change-map} and~\eqref{eq:relational1}, may be expanded in unhatted form as
\begin{align}
    {O}_{\tilde{T}_i|R'}&=\big[R_\star {T}_i\big]\circ R\circ R'^{-1}\\
    &=\tilde{T}_i\circ R'^{-1} = R'_\star\tilde{{T}}_i\,,\label{eq:scalarization}
\end{align}
where
\begin{equation}
    \tilde{{T}}_i\coloneqq\big[R_\star {T}_i\big]\circ R
\end{equation}
is a \emph{scalar} quantity on spacetime $\spacetime$ (for small diffeomorphisms), regardless of whether ${T}_i$ was. More precisely, if ${T}_i$ was a higher-rank tensor, then $\tilde{{T}}_i$ will be a set of scalars, one for each coefficient of ${T}_i$ (resp.\ $R_\star {T}_i$). Indeed, due to the small diffeomorphism \emph{invariance} of $R_\star {T}_i$ and \emph{covariance} of $R$, we have
\begin{equation}
    \tilde{{T}}_i[f_\star\phi]=\tilde{{T}}_i[\phi]\circ f^{-1}\,,\quad\,f\in\mathcal{G}\,.
\end{equation}
Only in the special case that ${T}_i$ is a scalar itself do we find the identity
\begin{equation}
    \tilde{{T}}_i={T}_i\,,\qquad {T}_i \text{ scalar}\,.\label{eq:scalarid}
\end{equation}

Accordingly, Eq.~\eqref{eq:scalarization} clarifies that ${O}_{\tilde{{T}}_i|R'}$ are simply the standard $R'$-frame dressings of the scalar gauge-covariant spacetime quantities $\tilde{{T}}_i$, i.e.\ the appropriate relational observables describing $\tilde{{T}}_i$ relative to $R'$. Note that, for any frame $R'$ that is fully independent of $R$, which in terms of their respective Maurer-Cartan forms and reorientations (see App.~\ref{app:geometry}) we can phrase as $\iota_{Y_{\rho_{R'}}}\omega_R=0=\iota_{Y_{\rho_R}}\omega_{R'}$, we have that ${O}_{\tilde{{T}}_i|R'}$ too is a set of scalars in relational spacetime $\ospace_{R'}$.\footnote{Clearly, this is not true for dependent frames $R,R'$, e.g.\ in the extreme case that $R'=R$, we reobtain ${O}_{\tilde{{T}}_i|R'}={O}_{{T}_i|R}$, which generally is a tensorial field in $\ospace_R$.} Eq.~\eqref{eq:tilderelational} implies that it is obtained by simply taking all the coefficients of $\hat{O}_{{T}_i|R}(o)$, treating them as scalars, and pulling them back to $\ospace_{R'}$ and evaluating at $o'=U_{R\to R'}(o)$. We call this a \emph{scalarization} of the $R$-relational observables in $R'$-relational spacetime. The fuzzy correlator in Eq.~\eqref{eq:fuzzycorrelator} is thus a correlator of these scalarized observables.

Next, let us specify how to think of the $\hat{O}_{\tilde{{T}}_i|R'}(o'_i[\hat\phi])$ (including their dependence on \emph{field-dependent} events in $\ospace_{R'}$) as functionals of the relational coordinates $\hat{O}^{A'}_{R'}$. This is relevant for understanding how to think of them inside the $n$-point functions and path integral in Eq.~\eqref{eq:fuzzycorrelator}. For simplicity, we focus on the example of scalar spacetime quantities ${T}_i$. The case of more general gauge-covariant quantities, such as higher-rank tensors, works analogously. 

Adapting the discussion in \cite[Sec.~3.3.4]{Goeller:2022rsx} to our case,\footnote{\cite[Sec.~3.3.4]{Goeller:2022rsx}, explains how to write relational observables, as quantities on $\ospace_{R}$, as composite single-integral observables on $\spacetime$ using the map $R[\phi]:\spacetime\to\ospace_R$ (denoted $R^{-1}[\phi]$ in \cite{Goeller:2022rsx}). Our discussion is the same upon replacing $\spacetime$ with $\ospace_{R'}$ and $R[\phi]$ with $U_{R'\to R}[\phi]:\ospace_{R'}\to\ospace_R$.\label{ftn:singleintegrals}} we have for a spacetime scalar quantity ${T}_i$
\begin{align}
    \hat{O}_{{T}_i|R}[\hat\phi](o_i)&=\hat{O}_{\tilde{{T}}_i|R'}(o'_i[\hat\phi])=\hat{O}_{{T}_i|R'}(o_i'[\hat\phi])\nonumber\\
    &=\hat{O}_{{T}_i|R'}[\hat\phi]\Big(\hat U_{R\to R'}[\hat\phi](o_i)\Big)\\
    &=\int_{\ospace_{R'}}{\rm{d}}^D o'\,\delta^{(D)}\left(o'-\hat{U}_{R\to R'}[\hat\phi](o_i)\right)\nonumber\\
    &\qquad\qquad\qquad\qquad\times\hat O_{{T}_i|R'}[\hat\phi](o')\,.\nonumber
\end{align}
Invoking the right expression in Eq.~\eqref{eq:frame-change-map}, we can rewrite the right hand side using
\begin{align}
    &\delta^{(D)}\left(o'-\hat{U}_{R\to R'}[\hat\phi](o_i)\right)\\
    &=\delta^{(D)}\left(\hat{O}_{R|R'}[\hat\phi](o')-o_i\right)\Bigg|\det\frac{\partial \hat{O}_{R|R'}[\hat\phi]}{\partial o'}\Bigg|\,\chi_{\hat{O}_{R|R'}[\hat\phi]}(o')\,,\nonumber
\end{align}
where $\chi_{\hat{O}_{R|R'}[\hat\phi]}$ is the characteristic function of the domain of $\hat{O}_{R|R'}=\hat{U}_{R'\to R}$ in $\ospace_{R'}$. In conjunction, this yields
\begin{align}
    &\hat{O}_{{T}_i|R}[\hat\phi](o_i)\label{eq:singleintegral}\\
    &=\int_{\ospace_{R'}}{\rm{d}}^D o'\,\delta^{(D)}\left(\hat{O}_{R|R'}[\hat\phi](o')-o_i\right)\Bigg|\det\frac{\partial \hat{O}_{R|R'}[\hat\phi]}{\partial o'}\Bigg|\nonumber\\
    &\qquad\qquad\qquad\quad\qquad\qquad\times\chi_{\hat{O}_{R|R'}[\hat\phi]}(o')\hat O_{{T}_i|R'}[\hat\phi](o')\,.\nonumber
\end{align}
The left hand side is a relational observable relative to $R$, evaluated at a field-independent event $o_i\in\ospace_R$, while the right hand side is a composition of relational observables delocalized via an integral relative to $R'$. Thus, for scalar quantities ${T}_i$, this provides the platform for extracting the functional relation $\hat{O}_{{T}_i|R}(o_i)\big[\hat O^{A'
}_{R'}\big]=\hat{O}_{\tilde{{T}}_i|R'}(o'_i[\hat\phi])\big[\hat O^{A
'}_{R'}\big]$, given that the $\hat{O}_{R|R'}$ and $\hat{O}_{{T}_i|R'}$ can be written in terms of them (or may even be among them). The precise functional form depends, of course, on the choice of relational coordinates $\hat{O}^{A'}_{R'}$, which is related to the choice of fields $\varphi^{A'}$ to coordinatize the sections $\mathcal{X}_{\gf{R}'}$. For non-scalar quantities ${T}_i$, one proceeds similarly by using  Eq.~\eqref{eq:scalarization}. 

The field-independence of $o_i\in\ospace_R$ has led, via the field-dependent map $\hat{U}_{R'\to R}$, to a completely delocalized integral observable over $\ospace_{R'}$. In other words, the \emph{field-dependent} localization of $\hat{O}_{{T}_i|R'}(o'_i[\hat\phi])$
in $\ospace_{R'}$ can be rewritten, using the delta function, as field-independent delocalization over $\ospace_{R'}$. This will be true also for more general relational observables: as a field on $\ospace_R$, $\hat{O}_{{T}_i|R}$ cannot have open $A'$-indices pertaining to relational spacetime $\ospace_{R'}$ of frame $R'$ (when the frames are independent). In other words, it can only feature  $A'$-indices whose $\ospace_{R'}$ part is \emph{contracted} and, recalling the properties of DeWitt notation, this means it will include one $\ospace_{R'}$-integral for each (partially) contracted pair. For example, in the above discussion of scalar observables, the field species part of $A'$ (the scalar) is left open, while the relational spacetime part is contracted.

We note that expressions as on the right hand side of Eq.~\eqref{eq:singleintegral} were termed `single-integral' observables in \cite{DeWitt:1962cg,Giddings:2005id,Marolf:2015jha,Marolf:1994wh}, with the important difference that in these references they were written as single integrals over spacetime $\spacetime$, cf.~footnote~\ref{ftn:singleintegrals}. Clearly, these are rather singular objects due to the delta function and would require regularization in the quantum theory, e.g.\ see \cite{Giddings:2005id} for some discussion on this. They may also easily be smeared in either $\ospace_{R}$ or $\ospace_{R'}$, see \cite[Sec.~3.3.4]{Goeller:2022rsx}.

In conclusion, in the identity of the relational correlators in Eqs.~\eqref{eq:relational-n-point} and~\eqref{eq:fuzzycorrelator},
\begin{align}
    &\left\langle
        \hat{O}_{{T}_1|R}(o_1)\cdots \hat{O}_{{T}_n|R}(o_n)
    \right\rangle_R\\
   &\qquad\qquad\qquad =   \left\langle
        \hat{O}_{\tilde{{T}}_1|R'}(o'_1[\hat\phi])\cdots \hat{O}_{\tilde{{T}}_n|R'}(o'_n[\hat\phi])
    \right\rangle_{R'} \,,\nonumber
\end{align}
we generally have a correlator of tensorial quantities in its sharp $\ospace_R$-representation on the left, and of scalar quantities in its fuzzy $\ospace_{R'}$-representation on the right. Of course, on the right, there is a scalar for each tensor component from the left; hence, if there is a tensor contraction on the left, there will be a corresponding sum of scalar correlators on the right. It is only when the ${T}_i$ are scalars that both sides correspond to correlators of scalar quantities, cf.~Eq.~\eqref{eq:scalarid}.

\section{Effective actions}

In this appendix, we discuss how our relational effective actions compare to previous proposals of effective actions, highlighting their conceptually and technically novel features. Subsequently, we also provide further detail on aspects of the relational effective action, especially commenting on quantum frame covariance.

\subsection{Comparison with previous proposals}
\label{app:EA_comparison}

Let us now compare our relational effective actions with previous  proposals, see Fig.~\ref{fig:effectiveactions} for a schematic summary.

\begin{figure*}[!t]
\centering
\scriptsize
\begin{tikzpicture}[node distance=2cm, scale=1.5]
  \draw[red!50, rounded corners=2mm, fill=red!5]
  (-4.5,-3.5) rectangle ++(4.5,3.7);
  \draw[blue!50!red!50, rounded corners=2mm, fill=blue!50!red!5] (-4.4,-3.4) rectangle ++(2.19,2.1);
  \draw[teal!50, rounded corners=2mm, fill=teal!5] (0.05,-3.5) rectangle ++(2.16,2.2);
  \node (dec1) [decision] {Gauge-fixing?};

\node (dec2a) [decision, below left of=dec1, yshift=-0.8cm, xshift=-1.9cm,] {Background gauge?};

\node (dec2b) [decision, below right of=dec1, yshift=-0.8cm, xshift=1.9cm,] {Field space covariance?};

\node (proc3a) [process, below left of=dec2a, yshift=-0.8cm, xshift=-0.7cm,] {Background field EA \cite{Parker:2009uva}};

\node (proc3b) [process, below right of=dec2a, yshift=-0.8cm, xshift=0.7cm,] {Dressed-fixed EA \cite{Falls:2025tid}};

\node (proc3c) [process, below left of=dec2b, yshift=-0.8cm, xshift=-0.7cm,] {Geometrical EA \cite{Vilkovisky:1984st,Burgess:1987zi,Huggins:1987zw,Lavrov:1988is}};

\node (proc3d) [process, below right of=dec2b, yshift=-0.8cm, xshift=0.7cm,] {Relational EA\\$\text{(our approach)}$};

\node (gaugeinv) [left of=dec1, xshift=-3.4cm, yshift=.45cm] {\textcolor{red!80}{Gauge dependence?}};
\node (background1) [left of=dec2a, xshift=-1.4cm, yshift=.7cm, anchor=west] {\textcolor{blue!50!red!80}{Background}};
\node (background2) [left of=dec2a, xshift=-1.4cm, yshift=.4cm, anchor=west] {\textcolor{blue!50!red!80}{dependence?}};
\node (comp) [left of=dec2b, xshift=-1.3cm, yshift=.4cm, anchor=west] {\textcolor{teal!80}{Computability?}};
\node (comp) [left of=dec2b, xshift=-1.3cm, yshift=.7cm, anchor=west] {\textcolor{teal!80}{Physical meaning?}};

\draw [arrow] (dec1) -| node[anchor=east] {yes} (dec2a);
\draw [arrow] (dec1) -| node[anchor=west] {no} (dec2b);

\draw [arrow] (dec2a) -| node[anchor=east] {yes} (proc3a);
\draw [arrow] (dec2a) -| node[anchor=west] {no} (proc3b);

\draw [arrow] (dec2b) -| node[anchor=east] {full} (proc3c);
\draw [arrow] (dec2b) -| node[anchor=west,text width=12mm] {frame changes; on shell} (proc3d);
\end{tikzpicture}
\caption{Different definitions of effective actions and their {challenges}. } 
\label{fig:effectiveactions}
\end{figure*}
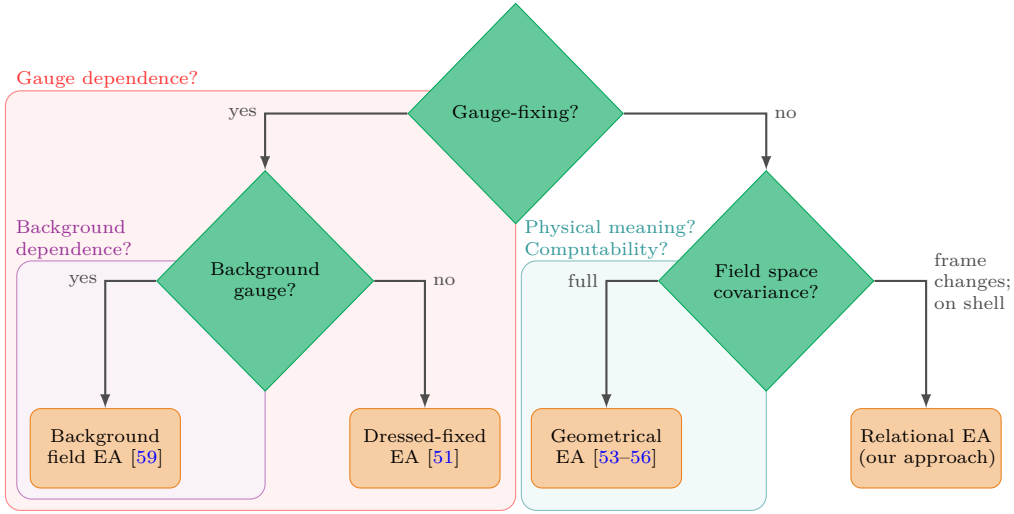
\paragraph*{Background-field effective action.}

The {background-field effective
action}~\cite{DeWitt:1967ub,DeWitt:1967yk,Abbott:1981ke,
Abbott:1980hw,Niedermaier:2006wt}, standard in continuum quantum gravity, splits the spacetime metric into a background and fluctuations,
$g_{\mu\nu}= \bar g_{\mu\nu}+ h_{\mu\nu}$, and gauge-fixes the
fluctuation relative to $\bar g$. With a background-covariant gauge fixing, the effective action is invariant under background gauge
transformations, while gauge transformations acting non-trivially on the fluctuations $h$, but trivially on the background metric, 
 are explicitly broken by the gauge fixing. Gauge
independence of physical observables holds perturbatively, but non-perturbative regularizations break it. The background--fluctuation split induces a bi-metric dependence, since the effective action generally depends separately on the background metric $\bar g_{\mu\nu}$ and the  full metric $g_{\mu\nu}=\bar g_{\mu\nu}+ h_{\mu\nu}$. This is encoded in modified Ward identities~\cite{Mottola:2003vx,Manrique:2009uh,
Becker:2014qya}. In contrast, our relational effective action requires neither a background--fluctuation split nor a (background) gauge fixing, as it is defined directly in terms of gauge-invariant relational observables. Nevertheless, such a split may still be introduced to define suitable regulators, as in the construction of a relational effective average action; see \cite{AFHM2}.
\paragraph*{Dressed-fixed effective action.}
The proposal of~\cite{Falls:2025tid} dresses the elementary fields using dynamical degrees of freedom whose dressing conditions are functionally equivalent to a gauge-fixing condition. Importantly, it absorbs the residual gauge breaking into the sources, which couple to the fundamental fields and auxiliary Nakanishi--Lautrup multipliers. On the Legendre-transformed side, gauge-equivalent configurations are still distinguished off-shell by
the sources. What is established is thus {covariance} of the effective action under a residual transformation, not manifest gauge invariance.
Equation~\eqref{eq:Z_relational} of the main text recovers this construction from our relational path integral upon fixing the frame configuration. However, crucially, our construction is manifestly gauge-invariant.
\paragraph*{Vilkovisky--DeWitt (`unique') effective action.}
The Vilkovisky--DeWitt construction
\cite{Vilkovisky:1984st,Burgess:1987zi,Huggins:1987zw,Lavrov:1988is} is based
on a ghost-free, fiber-adapted definition of the path integral
\cite{DeWitt:1980jv}, analogous to the one adopted in the main text. The main motivation for this approach was to formulate an effective action that is invariant under arbitrary field reparametrizations, i.e.\ to have field space diffeomorphism invariance. In this
case, however, the sources are not coupled linearly to the fields themselves,
but to a field space vector constructed from Synge's world function, associated
with a connection on field space \cite{Kunstatter:1990hz,DeWitt:1995cx,DeWitt:2003pm,gospel}. The resulting effective action is
gauge-invariant, and, by construction, invariant under arbitrary field reparametrizations.
This comes at a cost: it encodes information about $1$PI correlators of
field-space non-local objects. This makes its physical interpretation less
transparent, and renders concrete computations, especially in gravity, highly
challenging, for instance in renormalization or in applications of the LSZ
theorem \cite{DeWitt:2003pm}.

Our relational effective actions are not invariant under arbitrary field space diffeomorphisms, not even invariant under (quantum) \emph{off-shell} frame changes. However, it is  worth emphasizing that general field space covariance is not
required by any physical principle, since the field space geometry, including
its metric and connection, is non-dynamical. All that is required is that, for each frame perspective, we have a consistent effective action encoding correlators and fluctuations `as seen' by that frame. Since the relational field space coordinates are distinct for different frames, so are their correlators as we saw in the main text. Nevertheless, since the relational coordinates $\hat O^A_R$ are complete on patches of $\physf$, this should suffice to compute correlators in any frame perspective, as discussed shortly.

For these reasons,
we  propose a new formulation in terms of off-shell frame-dependent effective actions, arising naturally from a relational perspective, and
which generate $1$PI correlators of appropriate $\ospace_R$-\emph{local} relational observables,
while retaining full gauge and background independence. One may nevertheless require a weaker notion of on-shell covariance under frame changes, which can be seen as imposing conditions on the transformations relating frame-dependent sources, as explained below.

\subsection{The relational effective action and frame changes}
\label{app:relea}

The relational effective action was defined in Eq.\ \eqref{eq:relational-effective-action}. Equivalently, one may write
\begin{align}
    e^{i\Gamma_R[\bar{O}_R]}
    =
    \int \mathcal{D}
    \mu_R[\hat O_R]\,e^{i
        S[\hat O_R]
        -i\Gamma_{R,A}
        (
            \hat O_R^A
            -
            \bar{O}_R^A)}\,,
\label{eq:relational-ea-path-integral}
\end{align}
where $\Gamma_{R,A}\coloneq\delta\Gamma_R/\delta \bar{O}_R^A=-(J_R)_A$. 

Note that Eq.~\eqref{eq:relational-ea-path-integral} (much like \eqref{eq:relational-generating-functional} for $W_R$) fixes only $\exp(i\Gamma_R)$, and hence defines $\Gamma_R$ only up to shifts by $2\pi n_R$, with
$n_R\in\mathbb{Z}$. The integer $n_R$ may in principle depend on the choice of frame, but  is field-independent on each connected log-branch on which $\Gamma_R$ is a  differentiable functional: indeed, two  differentiable representatives whose exponentials agree can differ only by a continuous, integer-valued functional (times $2\pi$)
and, therefore, by a constant. This ambiguity is thus irrelevant for the quantum
equations of motion and for 1PI correlation functions, which depend on functional derivatives of $\Gamma_R$.

For each frame $R$, the Legendre relation
\begin{equation}
    \Gamma_{R,A}[\bar O_R] = -(J_{R})_A
\end{equation}
identifies the vanishing source locus $(J_R)_A=0$ with the quantum stationary locus $\Gamma_{R,A}=0$. The collection of these loci for all frames, therefore, provides a natural frame-perspective-neutral notion of quantum shell, which, for simplicity, will be referred to as the \emph{quantum shell} below. 

At this stage, we have not imposed any restriction on how the frame-dependent sources transform when changing frame. Frame covariance of the quantum equations of motion can then be seen as a physical condition on these transformations. It is a physically reasonable restriction, given that the common quantum solution space $J_R^A=0$ for all $A,R$ exists. Of course, for different $R$ these are solutions to distinct quantum equations of motion associated with distinct effective actions.

As argued in the main body, relational sources cannot be related by the usual frame-change maps $\hat U_{R\to R'}[\hat\phi]$, as these turn background fields into dynamical ones. This is in line with Eq.~\eqref{eq:frame-change} constituting \emph{active} frame transformations on the observables. Instead, what we will need here are the \emph{passive} transformations, which on the base $\physf$ yield a change of relational coordinates $\hat O^{R'}_{A'}[\hat O^B_R]$ (cf.~the discussion below Eq.~\eqref{eq:frame-change}). That is, we want a combined transformation for background sources and dynamical fields, $(\hat O^A_R,J_R)\to(\hat O^{A'}_{R'},J_{R'})$. Since the dynamical relational coordinates transform into one another and background fields must remain background, it follows that we can only have $J_{R'}[J_R]$, i.e.\ also the background sources must transform into one another. Glancing at Eq.~\eqref{eq:relational-ea-path-integral} then clarifies that $\Gamma_R$ cannot be off-shell covariant under such frame changes since, in general, the relation between $\hat O^A_R$ and $\hat O^{A'}_{R'}$ is highly non-linear (and similarly for their mean fields).  

However, requiring on-shell frame covariance restricts the source transformations to preserve the vanishing source loci, 
\begin{equation}
    J_{R'}[0] = 0 \, .
\end{equation}
This condition ensures that $\Gamma_{R,A}=0$ is mapped into $\Gamma_{R',A'}=0$, and hence that the quantum equations of motion are frame-covariant on-shell.

Now Eq.\ \eqref{eq:relational-ea-path-integral} implies that the value of the relational effective action on the $R$-quantum shell agrees with its value on the
$R'$-quantum shell, for any other frame $R'$, i.e.\ $\Gamma_R[\bar O_R]\approx\Gamma_{R'}[\bar O_{R'}]\approx -i\ln Z$ (up to the irrelevant multiples of $2\pi$), where `$\approx$' denotes equality on shell. In this sense, the relational effective action becomes a scalar under frame changes on the quantum shell. 
Of course, this concerns only the quantum shell, and does not require the relational effective action to transform as a scalar away from it.

It is useful to recast the above observation in infinitesimal terms by considering frame changes along a one-parameter 
family of frames $\{R_\lambda\}_{\lambda\in\mathbb{R}}$.\footnote{The space of all frames (or dressings) \cite{Goeller:2022rsx} is presumably not connected since there exist qualitatively quite distinct types of frames. For example, boundary anchored geodesic dressings yield frames that are built non-locally (in $\spacetime$) from the field content and lead to challenges to microcausality \cite{Donnelly:2015hta,Donnelly:2016rvo}, whereas frames locally constructed from matter do not lead to such challenges \cite{Marolf:2015jha,Goeller:2022rsx}. It is hard to see how one class may be deformed into the other in a continuous manner. For instance, not even all Lorentz tetrads are related by continuous transformations. Accordingly, the following infinitesimal analysis is only valid in each connected piece of the space of all frames.
}\footnote{For example, one may consider a family of relational frames defined by scalar fields that differ only in the value of their mass parameter. These frames may then be formally connected by continuously varying the mass, which plays the role of the parameter $\lambda$ here.} This can be done via a  
`frame-Nielsen identity', replacing the usual Nielsen identities, which encode how effective actions change under gauge flow \cite {Nielsen:1975fs,Fukuda:1975di}, or the modified split Ward
identities governing the dependence on the background-fluctuation split \cite{Pawlowski:2003sk,Pawlowski:2005xe,Manrique:2009uh,Donkin:2012ud}, by an identity encoding change under frame changes. The frame-Nielsen identity follows from varying both sides of Eq.\ \eqref{eq:relational-ea-path-integral} with respect to $\lambda$ (see \cite{AFHM2} for details):
\begin{equation}\label{eqn:framenielsen}
    \delta_\lambda(\Gamma_\lambda[\bar{O}_\lambda])=\Gamma_{\lambda,A}[\bar{O}_\lambda](\delta_\lambda \bar{O}^{A}_\lambda-\langle \delta_\lambda\hat{O}_{\lambda}^A\rangle_{J_\lambda})\,,
\end{equation} 
where $\Gamma_\lambda[\bar{O}_\lambda]\coloneq \Gamma_{R_\lambda}[\bar{O}_{R_\lambda}]$, $\delta_\lambda\coloneq \delta\lambda\diff/\diff\lambda$, {${\delta_\lambda \bar{O}^{A}_\lambda\coloneqq\delta_\lambda\langle \hat O^A_\lambda\rangle_{J_\lambda}}$, and $\langle\cdot\rangle_{J_R}$ denotes the \emph{off shell} mean value
\begin{equation}\label{eq:mean}
    \langle F[\hat{O}_R]\rangle_{J_R}\coloneq e^{-iW_R[J_R]}\langle e^{i[J_R]_A\hat{O}_R^A}\,F[\hat{O}_R]\rangle
\end{equation}
for any functional $F$ on $\physf$. On shell, $\Gamma_{\lambda,A}=0$, and hence $\delta_\lambda\Gamma_\lambda\approx 0$. This, together with the fact that $\delta_\lambda J_\lambda=0$ by definition of quantum shell  also implies that 
\begin{equation}\label{eq:transff}
    \delta_\lambda\langle {F}[\hat{O}_{\lambda}]\rangle_{J_\lambda}\approx \langle \delta_\lambda F[\hat{O}_{\lambda}]\rangle_{J_\lambda}\,,
\end{equation}
as one can see from Eq.\ \eqref{eq:mean}. In particular, it follows that $\delta_\lambda\bar{O}_\lambda^A\approx \langle \delta_\lambda\hat{O}_{\lambda}^A\rangle_{J_\lambda}$, which determines how the frame-dependent mean fields are related along the quantum shell. Note that Eq.\ \eqref{eq:transff} is conceptually and physically distinct from the transformations considered in Eq.\ \eqref{eqn:correlators}, which instead concern the rewriting of the same observables from the perspective of a different frame.

As argued above, the relational effective action does not, in general, transform as a scalar under frame changes (certainly not off-shell). A frame change may therefore be regarded as inducing a non-trivial change of the generating functional, and hence as a motion in theory space. Equivalently, such a motion can be parametrized in terms of changes of appropriate couplings \cite{Pagani:2017gnd}. By virtue of Eq.\ \eqref{eq:transff}, these transformations leave invariant the on-shell expectation values of frame-scalar quantities. Couplings that parametrize directions in theory space along which field space scalar observables remain invariant\,--\,typically because the corresponding transformations are generated by field redefinitions\,--\,are called \emph{inessential}, while the associated operators are called \emph{redundant} \cite{Wegner:1974sla,Dietz:2013sba, Baldazzi:2021ydj,Baldazzi:2021orb,Falls:2025sxu,Kuntz:2026vhs,Falls:2026nuh}. Infinitesimally, redundant deformations of the effective action are proportional to its equations of motion \cite{Baldazzi:2021ydj}. Frame changes therefore generate a class of inessential directions in theory space. In particular, the frame parameter $\lambda$ introduced above can be regarded as an inessential coupling, as Eq.\ \eqref{eqn:framenielsen} shows explicitly.

Finally, let us point out that Eq.\ \eqref{eqn:framenielsen} suggests a generalized notion of off-shell quantum frame covariance for the relational effective action. Combining Eq.\ \eqref{eqn:framenielsen} with
\begin{equation}
    \delta_\lambda\bigl(\Gamma_\lambda[\bar{O}_\lambda]\bigr)
    =
    \delta\lambda\,\partial_\lambda\Gamma_\lambda[\bar{O}_\lambda]
    +
    \delta_\lambda\bar{O}_\lambda^A
    \Gamma_{\lambda,A}[\bar{O}_\lambda]\,,
\end{equation}
we find that
\begin{equation}\label{eqn:generalizeddlambda}
    \nabla_\lambda\Gamma_\lambda[\bar{O}_\lambda]
    \coloneq
    \bigl(\partial_\lambda+\mathsf{L}_{\Xi_\lambda}\bigr)\,
    \Gamma_\lambda[\bar{O}_\lambda]
    =
    0
\end{equation}
off shell, where $\mathsf{L}_{\Xi_\lambda}$ denotes the field space Lie derivative along the vector field
\begin{equation}
    \Xi_\lambda
    \coloneq
    \Xi_\lambda^A\frac{\delta}{\delta\bar{O}_\lambda^A}\,,
    \qquad
    \Xi_\lambda^A
    \coloneq
    \left\langle
        \frac{\diff}{\diff\lambda}\hat{O}_\lambda^A
    \right\rangle_{J_\lambda}\,.
\end{equation}
Since both $\partial_\lambda$ and the Lie derivative commute with the field space differential, the above identity also implies
\begin{equation}
    \nabla_\lambda \Gamma_{\lambda,A}=0=\partial_\lambda\Gamma_{\lambda,A}+\Xi^B_\lambda\Gamma_{\lambda,AB}+\Xi^B_{,A}\Gamma_{\lambda,B}\,.
\end{equation}
This provides a modified, off-shell notion of frame covariance for the quantum equations of motion. On-shell, one has
$\Xi_\lambda^B\approx\diff\bar{O}_\lambda^B/\diff\lambda$ and
$\Gamma_{\lambda,B}=0$. The last equation then reduces to
$\delta_\lambda\Gamma_{\lambda,A}\approx0$, as expected.

The off-shell frame non-covariance of our relational effective actions notwithstanding, they may still be informationally complete in the following sense. While each $\Gamma_R$ directly encodes the information about sharp correlators in $R$-perspective via its functional derivatives, it would also indirectly encode the information about correlators (including fuzzy ones such as Eq.~\eqref{eq:fuzzycorrelator}) and expectation values in other frame perspectives in a joint domain, provided the sharp $n$-point correlators in $R$-perspectives are complete like the relational coordinates $\hat O^A_R$, which are complete on patches of $\physf$. By completeness of $n$-point correlators we mean that they would suffice to characterize the state (or pair of states in transition amplitudes). If they were complete (which arguably they should if the $\hat O^A_R$ are), then expectation values and correlators in other frame perspectives would be (in general highly non-polynomial) functions of the $n$-point functions in $R$-perspective.

\bibliography{references}

\end{document}